\documentclass[aps,prb,twocolumn,showpacs]{revtex4}
\usepackage{epsfig}

\begin{document}

\title{Positional Disorder, Spin-Orbit Coupling and Frustration in
${\rm Ga}_{1-x}{\rm Mn}_x{\rm As}$}
\author{Gregory A. Fiete$^{1-4}$, Gergely Zar\'and$^{1-3}$, Boldizs\'ar
Jank\'o$^{1,5}$, Pawel Redli\'nski,$^5$  and C. Pascu  Moca$^3$}

\affiliation{$^1$Materials Science Division, Argonne National
Laboratory, 9700 South Cass Avenue, Argonne, Illinois 60429, USA\\
$^2$Department of Physics, Harvard University, Cambridge,
Massachusetts 02138, USA\\ $^3$Research Institute of Physics,
Technical University Budapest, Budapest, H-1521, Hungary\\
$^4$Kavli Institute for Theoretical Physics, University of
California, Santa Barbara, CA 93106, USA\\ $^5$Department of
Physics, University of Notre Dame, Notre Dame, Indiana 46617, USA}

\date{\today}

\begin{abstract}

We study the magnetic properties of metallic ${\rm Ga}_{1-x}{\rm
Mn}_x{\rm As}$.  We calculate the effective RKKY interaction
between Mn spins using several realistic models for the valence
band structure of GaAs. We also study the effect of positional
disorder of the Mn on the magnetic properties. We find that the
interaction between two Mn spins is anisotropic due to spin-orbit
coupling both within the so-called spherical approximation and in
the more realistic six band model. The spherical approximation
strongly overestimates this anisotropy, especially for short
distances between Mn ions.  Using the obtained effective
Hamiltonian we carry out Monte Carlo simulations of finite and
zero temperature magnetization and find that, due to orientational
frustration of the spins, non-collinear states appear in both
valence band approximations for disordered, uncorrelated Mn
impurities in the small concentration regime.  Introducing
correlations among the substitutional Mn positions or increasing
the Mn concentration leads to an increase in the remnant
magnetization at zero temperature and an almost fully polarized
ferromagnetic state.

\end{abstract}
\maketitle


\section{Introduction}

Recently, there has been a surge of interest in the more than 30 year
old field of diluted magnetic semiconductors\cite{reviews} that has
been largely motivated by potential application of these materials in
spin-based computation\cite{divincenzo,nielson,Awschalom,Zutic} devices.  In
particular, the discovery of {\em ferromagnetism} in low-temperature
molecular beam epitaxy (MBE) grown ${\rm Ga}_{1-x}{\rm Mn}_x{\rm As}$
has generated renewed interest. In this material Curie temperatures as high as $T_c
\approx 160 {\rm K}$ have been observed.\cite{edmonds}

It is typically rather difficult to dissolve magnetic impurities
in a semiconductor using conventional MBE techniques, because the
magnetic ions usually tend to segregate from the host.  As a
result, these conventionally produced magnetic semiconductors
often exhibit spin glass physics at low temperatures and other,
undesirable, ``clustering'' effects.  Recently Ohno achieved the
key breakthrough,\cite{Ohno:apl96} by growing ${\rm Ga}_{1-x}{\rm
Mn}_x{\rm As}$ at close to room temperature, which minimized Mn
positional relaxation during the growth process, and resulted in a
material with intrinsic ferromagnetic properties, with a Curie
temperature $T_C\sim 110 {\rm K}$.

Post-growth annealing of the samples was shown to induce changes
both in the lattice and positional
defects\cite{potashnik,edmonds,yu,Hayashi:apl01} as well as in the
hole concentration.\cite{walukiewicz,Sorenson:apl03} Properties
such as magnetization and Curie temperature, were found to depend
very sensitively on details of the post-growth annealing
protocol\cite{ohno2,potashnik,edmonds2,Ku:apl03}.  In particular,
the measured $T=0$ temperature magnetization has been found to be
much less than one expected based on the nominal concentration of
Mn ions.\cite{Potashnik:apl03} The remnant magnetization and the
Curie temperature $T_C$ were both found to increase upon
annealing, but the magnetization reached only about half of the
expected value even in the case of optimally annealed samples
\cite{potashnik,Potashnik:apl03,Potashnik:prb02}. According to
recent experimental results, much of the missing magnetization is
probably due to interstitial Mn ions, which compensate valence
band holes,\cite{edmonds,yu} and, presumably, also bind
antiferromagnetically to substitutional Mn ions.\cite{bergqvist}
As a result, the nominal concentration of Mn ions, $x$, is usually
substantially larger than the concentration of {\em active} Mn ions
(participating in the ferromagnetism), $x_{\rm active}$, which is
larger than the concentration of mobile holes 
in the valence band (or possibly impurity band
\cite{fiete,berciu}), $c = x_{\rm active}\;f$, 
with $f<1$ the hole fraction. [In the
remainder of this paper - unless explicitly noted otherwise - we
present the results in terms of the active Mn concentration:
$x\equiv x_{\rm active}$.] However, especially for samples with a
lower doping level and/or shorter annealing times, even this
substantial difference in active and nominal Mn concentration
seems to be insufficient to explain the total amount of lost of
magnetization: In some of these samples the remnant magnetization
can be substantially increased by a relatively small magnetic
field, clearly hinting at a non-collinear magnetic
state.\cite{Potashnik:apl03}

Many of the properties of ${\rm Ga}_{1-x}{\rm Mn}_x{\rm As}$
depend crucially on defects, and understanding these defects is
currently a subject of intense
research.\cite{Timm:rev,edmonds,bergqvist,Blinowski:prb03,Korzhavi:prl02,yu,Maca:prb02}
${\rm Ga}_{1-x}{\rm Mn}_x{\rm As}$ is a complicated material:
Besides containing many defects of unknown origin, it is very
disordered, close to a localization transition with a mean free
path of the order of the Mn-Mn distance, has a complicated band
structure, has strong spin-orbit coupling and has a large exchange
coupling between the localized Mn spins and the valence holes.  It
is therefore almost impossible to study this material without
making some approximations.

As pointed out in Ref.~[\onlinecite{zarand}], spin-orbit coupling
can induce frustration and non-collinear states in a disordered
magnet. In this paper, we study the effect of spin-orbit coupling
in the metallic limit ({\em i.e.,} holes are assumed to reside
entirely in the valence bands of GaAs) of ${\rm Ga}_{1-x}{\rm
Mn}_x{\rm As}$, and study how it may influence the structure of
the ferromagnetic state. To this purpose, we will follow the route
outlined in Ref.~[\onlinecite{zarand}] and compute the effective
interaction between the spin of two substitutional Mn impurities
by doing perturbation theory in the exchange interaction (RKKY
interaction). It is important to note that other mechanisms also
exist that could lead to a non-collinear
state.\cite{schliemann,Schliemann:prl02} Recently, several studies
of RKKY models of Mn spin-spin interactions in
GaMnAs\cite{zarand,Priour:prl04,brey,Singh:prb03,Bouzerar:prb03,Timm:condmat04}
have been reported, but these studies have not investigated in detail
the effects of positional disorder on the magnetic properties
using realistic band structure models.

In this work, we neglect the effect of other defects and the strong
scattering potential on the charged substitutional Mn impurities.
This approximation - which is quite common in the
literature\cite{konig,abolfath} - is expected to provide {\em
qualitatively} reliable results 
relatively deep in the metallic state, far away from the
localization transition, where screening is
expected to reduce the effets of charged impurities. 
Quantitative agreement with experiment will require treating both 
charging effects and defect correlations in a realistic way.\cite{timm,Timm:rev} 
Furthermore, scattering from charged impurities becomes even more important 
in the dilute limit, and is ultimately responsible for the localization transition 
that occurs at small concentrations. In this dilute limit,  the strong 
potential scattering  off Mn impurities can be treated nonperturbatively, and 
results similar to those reported here are  obtained.\cite{fiete,Pawel}

In the absence of disorder the top of the valence band of ${\rm
Ga}{\rm As}$ can be described within the framework of $k\cdot p$
perturbation theory, which also accounts for the spin-orbit coupling
in this material, and gives a good description of the band structure
around the $\Gamma$ point. In particular, we study two different forms
of the $k\cdot p$ perturbation theory. First we study the effective
spin-spin interaction within a simplified version of the four band
model, the so-called spherical model,\cite{baldereschi} where only the
topmost four spin $j=3/2$ bands are kept and the distortion of the
spherical Fermi surface is neglected.  In this case analytical results
can be obtained for the effective Mn spin-spin interaction.\cite{zarand}  Then we
study the Mn spin-spin interaction using a six band
model\cite{abolfath,Konig:prb01} (Kohn-Luttinger Hamiltonian\cite{kohn}),
where the spin-orbit split $j=1/2$ band is also taken into account. In
the latter case it is not possible to evaluate the interaction kernel
analytically, and one must resort to numerics.

Although we find in both calculations a strong spin-orbit induced
anisotropy in the spin-spin interaction for typical Mn-Mn
distances, the spherical model largely overestimates the size of
the anisotropy for small distances. We find it especially
instructive to discuss this result in comparison with those
published recently in a nice and intriguing paper by Brey and
G\'omez-Santos. \cite{brey} While our result agrees qualitatively
with the one obtained in Ref.[\onlinecite{brey}], on a {\em
quantitative} level it is completely different. Brey and
G\'omez-Santos estimated that the largest value of the anisotropy
is of the order of $\sim 10^{-4}$. In contrast, in the present
calculation we find that the smallest value of the anisotropy is
around $1 \%$ for nearest neighbor Mn ions at a distance of $\sim
2.5 \AA$, and it increases continuously with distance, until it
reaches a value of the order of $\sim 20\; \%$ for typical Mn-Mn
distances.  These numbers roughly agree with the results obtained
in the very dilute limit for a ${\rm Mn}_2$ ``molecule'' within
the six band model.\cite{Pawel}

There are basically two reasons for the discrepancy between the
two results: \\ (1) Presumably to avoid convergence problems, Brey
and G\'omez-Santos introduced a short distance cut-off $a_0$, and
replaced the exchange interaction between the Mn core spins and
valence holes by a non-local interaction. The cut-off scheme they
introduced, however, is not compatible with the general structure
of the exchange interaction derived in the microscopic theory of
quantum impurities.\cite{Hewson} As we discuss in detail in
Sec.~\ref{sec:six_band},  the corresponding momentum cut-off
introduced should not be smaller than $\sim 2k_F$, the typical
momentum transfer during backscattering.  Unfortunately, the
cut-off $a_0 = 4 \AA$ used in Ref.~[\onlinecite{brey}] did not
satisfy this criterion, and as a consequence, the results of
Ref.~[\onlinecite{brey}] showed a very strong dependence on $a_0$:
In particular, for somewhat smaller values of $a_0 = 2.42 \AA $
Brey and G\'omez-Santos found an anisotropy of a few percent,
which roughly agrees with the one we find ($\approx$ 1 \%) within a
cut-off scheme compatible with the general theory of magnetic
interactions for nearest neighbor Mn ions. The strong dependence
of the Brey-G\'omez-Santos result on the cut-off parameter was
also pointed out and questioned recently by Timm and MacDonald in
Ref.[\onlinecite{Timm:condmat04}].
We emphasize that with our cut-off scheme the anisotropy depends only {\em weakly} on the value of $a_0$.  \\
(2) Secondly, in Ref.~[\onlinecite{brey}] it has been assumed that
the anisotropy is largest for the shortest Mn-Mn separations.
However, for hole concentrations $p\le 3 \ {\rm nm}^{-3}$ one
actually expects on general grounds that the {\em asymptotic} form
of the RKKY interaction is reasonably well-approximated within the
four band model, which assumes infinite spin-orbit splitting
$\Delta_{SO}$, and that the anisotropy increases with Mn-Mn
separation. This is indeed supported by our numerical results: For
the very small separations for which Brey and G'omez-Santos
carried out their calculations we indeed find only a $\sim 1 \%$
anisotropy, while for {\em typical} Mn-Mn separations we find a
rather large anisotropy of the order of $\sim 20\%$.

Having the interaction kernels at hand, we considered a
distribution of Mn ions and carried out finite temperature Monte
Carlo simulations to characterize the magnetic properties of the
system.  At present, it is unclear from the experiments to what
extent the positions of {\em substitutional}, magnetically active
Mn are correlated during the growth process.  In order to gain
insight into how Mn positional correlations may affect the
magnetic properties of a material like GaMnAs, we studied the
dependence of the Curie temperature, the saturation magnetization,
the shape of the magnetization curve and the ground state spin
distribution on Mn positional correlations.  These correlations
were introduced by allowing repulsive interactions between the Mn
ions and relaxing them through a zero temperature Monte Carlo
Metropolis algorithm in a way similar to Ref.~[\onlinecite{timm}].
We started the simulations from a set of completely uncorrelated
initial Mn positions, and allowed the substitutional Mn ions to
move.  We then studied the magnetic properties of the spin system
as a function of the Mn positional relaxation time, including the
completely uncorrelated configuration.

Our main results are as follows: Within the spherical approximation we
find a strongly disordered (non-collinear) ferromagnetic ground state
for unrelaxed Mn positions, where the ground state magnetization can
be reduced by as much as $\sim 50 \%$ with respect to its saturation
value. Upon {\rm relaxing} the Mn positions, however, the remnant
magnetization increases, and for a fully relaxed lattice one may
recover as much as $95\%$ of the saturation value. In contrast to our
earlier expectations,\cite{zarand} the obtained non-collinear state is
{\em not} a spin-glass: Although a finite system displays hysteresis,
the coercive field we compute {\em decreases} with system size,
implying that the hysteresis vanishes in the infinite system size
limit.  The frustration effects found within the spherical model are
very sensitive to the hole fraction $f$ characterizing the degree of
compensation, and become more pronounced for larger hole fractions.

Spin-orbit coupling induced anisotropy does not have such a
dramatic effect if we use the kernel obtained within the six-band
model.  For a substitutional Mn concentration of $x=3\%$, hole
fraction $f=0.4$, and unrelaxed Mn spins we find that the
magnetization is reduced by $\sim 5 \%$, relative to the fully
polarized state, corresponding to a typical non-collinear angle of
the order of $\theta \sim 16$ degrees. This non-collinearity is,
however, almost completely suppressed if we let the Mn positions
relax using the Monte Carlo algorithm mentioned above. By
increasing the substitutional Mn concentration to $x=5 \%$ the
ground state becomes almost fully collinear even without relaxing
the Mn positions. The results do not depend too strongly on the
level of compensation (hole fraction). Our results also indicate
that anisotropy effects may be more important for smaller Mn
concentrations.\cite{fiete}

We emphasize again that the concentration $x$ above denotes the
concentration of {\em active} Mn ions, {\em i.e.} the
concentration of those Mn ions that participate in the formation
of the ferromagnetic state. Due to compensation effects induced by
interstitial Mn impurities, $x$ is usually much less than the
nominal concentration of Mn ions. In fact, samples with a nominal
concentration of $5 \%$ may easily have an active Mn concentration
in the range $x < 0.02\;-\;0.03$. In this concentration range the
impurity band approach of Refs.~[\onlinecite{berciu}] and
[\onlinecite{fiete}] may be more appropriate.

In agreement with the results of Ref.~[\onlinecite{Berciu:prb04}], we
also find that in all cases $T_C$ decreases as we relax the Mn
impurities, irrespective of the specific form of effective interaction
used, and $M(T)$ becomes more mean-field like at the same time.

This paper is organized as follows. In Sec.~\ref{sec:interaction}
we present the RKKY calculation within the Baldereschi-Lipari
spherical approximation for the effective interaction between two
Mn spins. Several of the most technical details and lengthy
expressions have been relegated to
Appendices~\ref{app:RKKY_calculation} and \ref{app:appendix_int}.
Sec.~\ref{sec:six_band} is devoted to the computation of the
effective Mn spin-spin interaction within the full six band model.
In Sec.~\ref{sec:montecarlo} we discuss the results of classical
Monte Carlo calculations for the magnetization, the ground state
spin distributions, the Curie temperature and the dependence of
these quantities on positional disorder and hole concentration
using the effective spin-spin interactions computed in
Sections~\ref{sec:interaction} and \ref{sec:six_band}.  Finally,
in Sec.~\ref{sec:conclusions} we summarize our main conclusions.

\section{Effective Mn Spin-Spin Interaction in the Spherical Approximation}
\label{sec:interaction}

In the present section, we study the effective RKKY interaction
between two Mn impurities, by using the so-called spherical
approximation to describe the valence band hole fluid. The top of
the valence band of GaAs consists of six $p$-bands. Four of these
bands become fourfold degenerate at the $\Gamma$ point and have
spin $j=3/2$ character, while the other two are split off by
$\Delta_{SO} \approx 0.34\; {\rm eV}$ and have spin $j=1/2$
character.\cite{Blakemore:jap82}

The spherical approximation consists of setting the spin-orbit splitting
$\Delta_{SO}$ to infinity, keeping only the
spin $j=3/2$ bands, neglecting the warping of the
Fermi surface due to the cubic symmetry of the crystal, and
approximating the dispersion of the holes by the following
expression:\cite{baldereschi}
\begin{equation}
 H_{\rm sp}={\gamma_1 \over 2 m} \left (p^2 -\nu \sum_{\alpha, \beta} J_{\alpha \beta} p_{\alpha \beta} \right )\;.
\label{eq:H_sp}
\end{equation}
 Here  $J_{\alpha \beta}={1 \over 2} (j_{\alpha}j_{\beta}+
j_{\beta} j_{\alpha} )-{1\over 3} \delta_{\alpha \beta} {\rm
Tr}\{j_{\alpha} j_{\beta} \}$ is the quadrupolar angular momentum
of the holes. The  linear momentum tensor of the holes $p_{\alpha
\beta}$ is defined  likewise. The coupling $\nu =(6 \gamma_3 + 4
\gamma_2)/5 \gamma_1\approx 0.77$ characterizes the strength of
the effective spin-orbit coupling within the $j=3/2$ band, and the
$\gamma_i$'s are the so-called Luttinger parameters\cite{kohn},
characterizing  the band structure of GaAs. Physically the
parameter $\nu$  describes how the spin of a $j=3/2$ valence hole
couples to its orbital motion.

The advantage of the spherical approximation, Eq.~(\ref{eq:H_sp}), is
that it makes it possible to obtain analytical results while it gives
a realistic description of the top of the valence band.  As we will
see, this latter fact is somewhat misleading: While providing a
qualitatively correct description of the valence bands of
Ga$_{1-x}$Mn$_x$As, it gives quantitatively incorrect results for the
RKKY interaction between two local spins.

In GaMnAs subsitutional Mn ions are believed to be in a ${\rm
Mn}^{2+}$ configuraration\cite{linnarson,szczytko,okabayashi_0}
thereby behaving as a negatively charged scatterer and having a core
spin $S=5/2$.  This core spin then couples to the spin of the valence
hole fluid through an exchange interaction that, for a single Mn ion,
 takes the following form within the spherical approximation:
\begin{equation}
H_{\rm int}({\bf R})=G {\bf S \cdot j(R)}\;,
\label{eq:exchange}
\end{equation}
where ${\bf j(R)}$ denotes the  valence holes' spin density, at the
location of the Mn ion, ${\bf R}$, and  $G>0$ is an antiferromagnetic
coupling.

To proceed let us first show that the eigenstates of $H_{\rm sp}$
are chiral, {\em i.e.}, that the spin is quantized along the
direction of propagation. First we note that $H_{\rm sp}$ for
holes propagating along the $z-$direction is easily  diagonalized:
${\bf p}=(0,0,p_z)$ so that
\begin{eqnarray}
\sum_{\alpha, \beta} J_{\alpha \beta} p_{\alpha \beta} &=& \left (j_z^2 -\frac{j(j+1)}{3} \right )p_z^2
\nonumber = \pm p_z^2\;.
\end{eqnarray}
Thus, for holes moving in the $z-$direction,
\begin{equation}
H_{\rm sp} = {\gamma_1 p_z^2\over 2 m}(1\pm \nu)\;,
\label{eq:H_z_move}
\end{equation}
with the plus and minus signs standing for $j_z=\pm 1/2$ and
$j_z=\pm 3/2$, respectively.
This defines the light hole mass, $m_l=\frac{m}{\gamma_1 (1+\nu)}\approx
0.07 m$ and the heavy hole mass $m_h=\frac{m}{\gamma_1
(1-\nu)}\approx 0.5 m$ where $m$ is the bare electron mass.
For holes propagating in a general direction, the wave function can be
constructed  by spin 3/2 rotations so that
$\mu = {\bf j}\cdot {\bf \hat k}=\pm3/2$ for
heavy holes and $\pm 1/2$ for light holes propagating along direction
$\hat {\bf k}$.

The kinetic energy $H_{\rm sp}$ greatly simplifies in the  basis of these
chiral states, and takes the following form in second quantized formalism:
\begin{equation}
\hat H_{\rm sp} = \sum_{ {\bf k}, \mu} {k^2 \over 2 m_{\mu}} c^\dagger_{{\bf k} \mu} c_{{\bf k} \mu} \;.
\label{eq:H_chiral}
\end{equation}
Here $c^\dagger_{{\bf k} \mu}$ denotes the creation operator of a
hole with wavevector ${\bf k}$ and spin projection $\mu$ along $\hat
{\bf k}$.  Diagonalizing $\hat H_{\rm sp}$, however, produces a
strongly momentum dependent carrier-impurity interaction:
\begin{equation}
\hat H_{\rm int}={G\over V} \sum_{ {\bf k}, {\bf k'}}\sum_{\alpha,\mu,\mu'} S^{\alpha} c^\dagger_{{\bf k} \mu} j^{\alpha}_{\mu \mu'}(\hat {\bf k}, \hat {\bf k'})c_{{\bf k'} \mu'}
e^{-i{\bf (k -k')R}}\;.
\label{eq:Hint}
\end{equation}
Here $j^{\alpha}_{\mu \mu'}(\hat {\bf k}, \hat {\bf k'}) \equiv
\sum_{j,j'} D^\dagger_{\mu j}(\hat {\bf k})j^{\alpha}_{j j'} D_{j'
\mu'}(\hat {\bf k'})$ where $D(\hat {\bf k})$ is the spin 3/2 rotation
matrix and V is the volume of the sample. The rotation matricies are
defined here in the usual way as
$D_{j \mu}(\hat {\bf k}) = [e^{-i {\hat j^z \phi \over
\hbar}}e^{-i {\hat j^y \theta \over \hbar}}]_{j \mu}$,
where $\phi$ and $\theta$
are the usual azimuthal and polar angle of the direction ${\bf \hat
k} = (\sin \theta \cos \phi, \;\sin \theta \sin \phi,\; \cos \theta)$.

Having diagonalized $H_{\rm sp}$, we now consider two Mn ions at
positions ${\bf R}$ and the origin,  do perturbation  theory in
the exchange coupling $G$ and compute the RKKY interaction between
two Mn spins.\cite{yosida} The first order correction to the
single particle eigenstate $|{\bf k}, \mu\rangle_0$  of  $H_{\rm
sp}$ is given by
\begin{equation}
|{\bf k}, \mu \rangle'=|{\bf k}, \mu \rangle_0
-{G \over V} \sum_{\alpha, {\bf k'},\mu'}
{j^{\alpha}_{\mu \mu'}(\hat {\bf k}, \hat {\bf k'})
S^{\alpha}  \over \epsilon_{{\bf k'} \mu'} - \epsilon_{{\bf k} \mu} }
|{\bf k'}, \mu' \rangle_0\;,
\end{equation}
where the sum is over all  states except for  $|{\bf k}, \mu\rangle_0$,
and $\epsilon_{{\bf k} \mu} = k^2/2m_\mu $.
With these single particle states in hand, we can then compute the
shift of the ground state energy as:
\begin{eqnarray}
H_{\rm RKKY}&=&- 2\left ({G \over V}\right )^2 \sum_{\alpha \beta}{\bf S_R^{\alpha} S_0^{\beta}}\nonumber \\
&& \times \sum_{\scriptstyle {\bf k},\mu \atop \scriptstyle {\bf k'} \mu'}
{1 \over \epsilon_{{\bf k'} \mu'} - \epsilon_{{\bf k} \mu} }
\left(e^{i {\bf (k'-k)R}} j^{\beta}_{\mu \mu'} j^{\alpha}_{\mu' \mu}
+ c.c.\right )\;, \nonumber \\
\label{eq:RKKY_sum}
\end{eqnarray}
where we have dropped a piece of the energy independent of relative spin
orientations. This equation can be rewritten as
\begin{equation}
H_{\rm RKKY}=-\sum_{\alpha \beta} S^{\alpha}({\bf R})K^{\alpha \beta}({\bf R}) S^{\beta}(0)\;,
\end{equation}
with an obvious definition of  the kernel $K^{\alpha \beta}$.
Within the spherical approximation this result  further simplifies
to
\begin{equation}
H_{\rm eff}= -K_{\rm para}(R) {\bf S}_1^{||}\cdot {\bf S}_2^{||}-K_{\rm perp}(R) {\bf S}_1^{\perp}\cdot {\bf S}_2^{\perp}\;,
\label{eq:Heff}
\end{equation}
where $R=|{\bf R}_1-{\bf R}_2|$ and ${\bf S}_i$ is the Mn spin at position
${\bf R} _i$. Here ${\bf S}^{\perp/||}$ refers to the component of the spin
perpendicular/parallel to the axis joining them.

The computation of the kernels $K_{||}$ and $K_{\perp}$ is not straightforward
even in the  case of the spherical approximation.
Their explicit expressions  and some details about how to
obtain them are  given in Appendix~\ref{app:RKKY_calculation}.
These expressions together with Eq.~(\ref{eq:Heff}) constitute the main
result of this section of the paper.

\begin{figure}[h]
\begin{center}
\epsfig{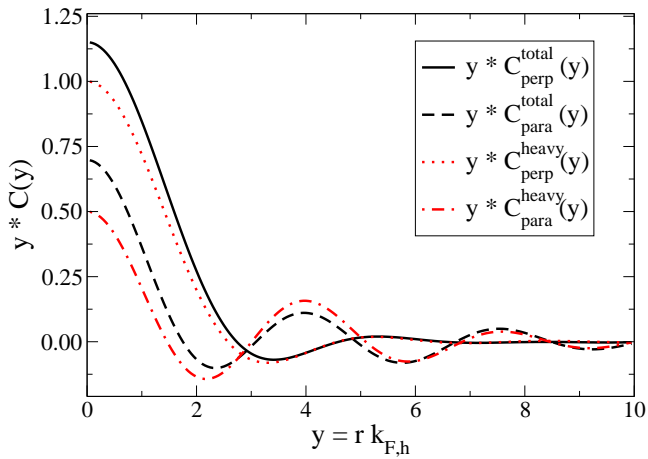}
\caption{
Effective RKKY kernel in the 4-band spherical
approximation. The contribution from the heavy hole
sector is also shown.  At typical Mn-Mn separations
the structure of the kernel is very sensitive to the value of
the hole fraction $f$.
For  $f=0.10$ one has $y^{\rm typ} = d^{\rm typ} k_{F,h}=1.6$, while
for  $f=0.25$   we have $ y^{\rm typ} = d^{\rm typ} k_{F,h}=2.2$.
In both cases the perpendicular part of the kernel is ferromagnetic,
but for $f=0.25$ the parallel component is antiferromagnetic and
is comparable in absolute value to the perpendicular component.
\label{fig:reduced_kernel}}
\end{center}
\end{figure}

For more physical transparency, it is worth defining the following
rescaled  kernel,
\begin{equation}
C_{\rm para/perp} \equiv  K_{\rm par/perp} / 4 \pi\,\epsilon_F g_{h}^2\;,
\end{equation}
where we have defined the dimensionless
coupling as $g_{h}=G \varrho_{h}=G {m_h^{3/2} \over \sqrt{2}
\pi^2}\epsilon_F^{1/2}$,  with
$m_h$ is the heavy hole mass and $\epsilon_F$ is the Fermi energy measured
relative to the ``top'' of the
valence band.   The positional dependence of $C_{\rm para/perp}$
is shown in Fig.~\ref{fig:reduced_kernel}, where we also plotted the
contribution of the
heavy holes, giving the dominant contribution to the interaction kernel.
The rather involved explicit expressions for
the dimensionless kernel $C(y)$
are given in Appendix~\ref{app:RKKY_calculation}.
The asymptotic forms for $y\equiv k_{F,h}R\to 0$ of the kernels $C$
are given by
\begin{eqnarray}
C_{\rm para}^{\rm total} \approx \left [
1 + \frac{m_l^{3/2}}{m_h^{3/2}}
+ \frac{7}{3} \left(\frac{m_l}{m_h}
+ \frac{m_l^{5/2}}{m_h^{5/2}}\right)
\right]
\frac{1}{2y}\nonumber
=\frac{.698}{y}\;,
\end{eqnarray}
and
\begin{eqnarray}
C_{\rm para}^{\rm total} \approx \left [
1 + \frac{m_l^{3/2}}{m_h^{3/2}}
+ \frac{2}{3} \left(\frac{m_l}{m_h}
+ \frac{m_l^{5/2}}{m_h^{5/2}}\right)
\right]
\frac{1}{y}\nonumber
=\frac{1.150}{y}\;,
\end{eqnarray}
where $m_l$ and $m_h$ are the light and heavy hole mass defined just
below Eq.~(\ref{eq:H_z_move}).
Specifically, in the limit of $m_h \to \infty$ we simply obtain
$C_{\rm para}^{\rm heavy} \approx 1/2y$,
$C_{\rm perp}^{\rm heavy} \approx 1/y$.

Eq.~(\ref{eq:Heff}), was first obtained in Ref.~[\onlinecite{zarand}]
using only the heavy hole sector of the valence band structure. There
it was shown that Eq.~(\ref{eq:Heff}) leads to a spin-disordered
ground state whose signatures seem to be present  in
some experiments.\cite{Potashnik:apl03,Rappoport:prb04}

At typical Mn-Mn separations the kernel is very sensitive to the value
of the hole fraction. Hole fractions in the range $f=0.10-0.25$
corresponds to ${\rm y^{typ}}_{f=0.10}={\rm d^{typ}}
k^{f=0.10}_{F,h}$=1.6 and ${\rm y^{typ}}_{f=0.25}={\rm d^{typ}}
k^{f=0.25}_{F,h}$=2.2.  In both cases the perpendicular part of the
kernel is largest and is ferromagnetic.  However, for $f=0.10$ the
parallel component is ferromagnetic while for $f=0.25$ the parallel
component is antiferromagnetic and of comparable strength to the
perpendicular component.
The relative size of $K_{\rm perp}$ and $K_{\rm para}$ means that spins
prefer to ferromagnetically align perpendicular to the axis joining
them for small enough hole concentrations.  In general this leads to
frustration and non-collinear magnetism.

In Sec.~\ref{sec:montecarlo} we study the finite temperature
behavior of the Mn spins using classical Monte Carlo techniques. There we
shall assume that the spins behave as classical objects, and we
replace Eq.~(\ref{eq:Heff}) by
\begin{equation}
H_{\rm eff}= -J_{||}(R) {\bf \Omega}_1^{||}\cdot {\bf \Omega}_2^{||}-J_{\perp}(R)
{\bf \Omega}_1^{\perp}\cdot {\bf \Omega}_2^{\perp}\;,
\label{eq:Heff_2}
\end{equation}
where now ${\bf \Omega}_1$ and ${\bf \Omega}_2$ denote the two unit vectors specifying
the direction of the spins, and the exchange couplings are given by
\begin{equation}
J_{ || / \perp} = E_J \;C_{\rm para/perp}\;,
\end{equation}
with the exchange energy defined as
\begin{equation}
E_J \equiv S^2 \;4 \pi\,\epsilon_F g_{h}^2\;.
\end{equation}
The precise value of $E_J$ depends on the specific value of the
unknown coupling
$G$  and the hole concentration, $p$, with which it scales as
$E_J\sim \epsilon_F^2\sim p^{4/3}$  within the spherical approximation.

\section{Effective Mn spin-spin interaction in the six band model}
\label{sec:six_band}

In the previous section we studied the effective RKKY interaction
between two Mn impurities within the spherical approximation. In
the present section we shall study the effective Mn spin-spin interaction
by using a more realistic description of the top of the valence band,
the so-called six band model, where the Hamiltonian reads
\begin{eqnarray}
H &=& H_{\rm Lutt} + H_{\rm int}\;,\\
H_{\rm int} &=& J_{pd} ({\bf s}({\bf R}_1)\cdot {\bf S}_1 +    {\bf s}({\bf R}_2)\cdot {\bf S}_2)\;,
\label{eq:basic}
\end{eqnarray}
with $H_{\rm Lutt}$ being the so-called Kohn-Luttinger Hamiltonian, and
${\bf s}({\bf R})$ the spin density of the holes at position ${\bf
R}$. Note that ${\bf s}$ here is {\em not} a spin $3/2$ spin
object. It is a $6 \times 6$ dimensional matrix that represents
the spins (1/2) of the holes in the three $p$ channels that
constitute the top of the valence band (for an explicit definition
see {\em e.g.} Ref.~[\onlinecite{abolfath}]).

The six-band Hamiltonian approximates rather well the band structure
and the spin content of the hole states in the vicinity of the
$\Gamma$ point of the Brillouin zone. However, it is impossible to
evaluate the RKKY interaction analytically using $H_{\rm Lutt}$, and
numerical calculations are needed.

To perform the numerical calculations, we followed the procedure
of Ref.~[\onlinecite{brey}]: We generated a mesh in
momentum-space, computed the eigenenergies of $H_{\rm Lutt}$ and
the corresponding eigenfunctions, and ordered them according to
their energy. We then computed the matrix elements of $H_{\rm
int}$ in this basis by treating the spin of the Mn ions
classically, and fixing their direction. Next we diagonalized
$H_{\rm Lutt} + H_{\rm int}$ to obtain the single particle hole
states for this orientation of the spins, while only keeping
states below the cut-off energy, $E_{\rm cutoff}$. Finally, we
computed the ground state energy of the whole system as a sum of
the energy of the occupied single particle states, and thus
obtained the ground state energy of the system for a {\em fixed}
number of holes as a function of the direction of the two spins.

The effective couplings have been extracted from the ground state
energy in the same way as in Ref.~[\onlinecite{brey}]: First we placed
the two Mn ions in a distance $R$ along the $z$-direction and computed
the ground state energies for a fixed number of holes in a case where
the two spins were (1) parallel and pointing along the $z$-direction
($E_{\uparrow\uparrow}$), (2) oppositely aligned along the
$z$-direction ($E_{\uparrow\downarrow}$), and (3) parallel along the
$x$ direction ($E_{\rightarrow\rightarrow}$). In this arrangement the
effective coupling and anisotropy can be defined as:
\begin{eqnarray}
&& 2J_{||} \equiv E_{\uparrow\downarrow} -  E_{\uparrow\uparrow} \;,
\\
&& \delta J \equiv J_{||} - J_\perp \equiv E_{\rightarrow\rightarrow} -  E_{\uparrow\uparrow} \;.
\end{eqnarray}

\begin{figure}
\epsfxsize=2.5in
\begin{center}
\includegraphics[width=8.5cm,clip]{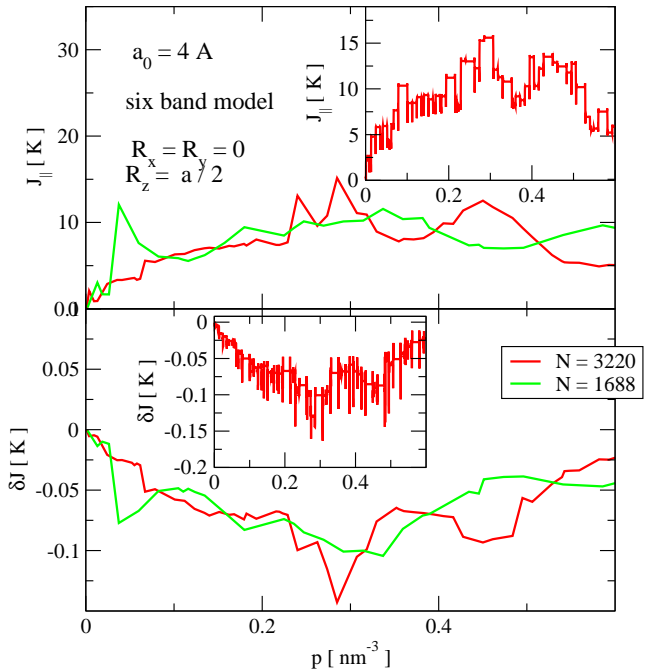}
\end{center}
\vspace*{0.05cm} \caption{ Effective exchange $J$ and anisotropy
$\delta J$ for two Mn ions. We used the cut-off scheme
(\ref{eq:good_cut-off}) with $a_0 = 4 \;\AA$, and a cut-off energy
$E_{\rm cutoff} = 3273\; {\rm K}$. The main parts show
the obtained interaction for $L = 23 a$ and $L=29 a$
at concentrations corresponding to ``closed shells'', while the insets
show the $L=29 a$  results for all concentrations.}
\label{fig:mesh_dep}
\end{figure}

There are, however, a few technical details that should be
discussed. First, the $k$-points in Brillouin space must be
generated in such a way that they respect the cubic symmetry of
the crystal, otherwise the obtained effective interaction breaks
the cubic symmetry. Due to the cubic symmetry, however, the
electronic levels typically become extremely degenerate for
$J_{pd} = 0$ and for a typical mesh of $k$ points this degeneracy
can reach numbers as high as $96$ in a mesh size of 1000. One
therefore experiences {\em large} but apparently systematic
fluctuations of the exchange energy computed as one fills the
single particle energy levels for $J_{pd} \ne 0$ one by one (see
the inset of Fig.~\ref{fig:mesh_dep}), and results converge very
slowly with increasing mesh size.  There are, however, special
points where the number of holes is such that for $J_{pd}=0$ an
integer number of degenerate energy ``shells'' are occupied.  At
these points the calculations seem to converge somewhat faster,
and an accuracy as high as $10-20\%$ can be reached. This
technique has been exploited in Ref.~[\onlinecite{brey}], and we
will also use it to compute the effective interaction kernel.

Another very important issue is the cut-off scheme used: To facilitate
convergence, it is important to introduce a cut-off for the exchange
coupling in momentum space.  Brey and G\'omez-Santos therefore replaced
$J_{pd}$ in Eq.~(\ref{eq:basic}) by a non-local interaction
between the Mn spin and the valence hole,
$J_{pd}
\to J_{pd}({\bf r} - {\bf R})$, and introduced a corresponding
cut-off for the exchange interaction in k-space of the form:
\begin{equation}
J_{pd}({\bf k} , {\bf k}') = J_{pd} e^{-({\bf k} - {\bf k}')^2 a_0^2/2}\;,
\label{eq:wrong_cut-off}
\end{equation}
where $a_0$ is the range of the non-local interaction.

Unfortunately, there is a serious problem with the cut-off
scheme (\ref{eq:wrong_cut-off}) for large values of $a_0$:
Eq.~(\ref{eq:wrong_cut-off}) gives a cut-off for the {\em momentum
transfer}.  On the other hand, it is well-known from the Kondo
literature that the exchange interaction deduced from the more
fundamental Anderson impurity problem has the following
structure:\cite{Hewson}
\begin{equation}
J_{pd}({\bf k} , {\bf k}') \sim V_{\bf k}V^\star_{{\bf k}'}
\left( {1 \over U + \epsilon_d} - {1\over \epsilon_d} \right)\;,
\label{eq:good_cutoff}
\end{equation}
where $\epsilon_d<0$ is the energy of the d-level, $U$ is the
Hubbard interaction, and the $V_{\bf k}$'s denote the ${\bf k}$-dependent
hybridization with states within the Brillouin zone.
The $V_{\bf k}$'s do have a momentum cut-off for momenta of the
order of the size of the Brillouin zone, however, $J({\bf k} ,
{\bf k}')$ clearly factorizes, and there is obviously {\em no cut-off for the
momentum transfer across the Fermi surface}.\cite{footnote2}

Since back scattering with large momentum transfer is responsible
for much of the leading term in the asymptotic RKKY interaction,
it is essential to use a cut-off scheme which does not influence
these large momentum transfer electron-hole excitations in the
vicinity of the Fermi surface, and is consistent with
Eq.~(\ref{eq:good_cutoff}).  In fact, the value $a_0=4\AA$ chosen
by Brey and G\'omez-Santos seems to be too large: They find that
the relative strength of the anisotropy is very sensitive to the
precise value of $a_0$, and increases dramatically as they
decrease it to $a_0 = 2.5\AA$, suggesting that the results
obtained by Brey and G\'omez-Santos with $a_0 = 4\AA$ are not
reliable.

To avoid this problem we therefore used the following cut-off scheme,
\begin{equation}
J_{pd}({\bf k} , {\bf k}') = J_{pd} \; e^{-({\bf k}^2 +  {\bf k}'^2) a_0^2/2}
\;,
\label{eq:good_cut-off}
\end{equation}
which is consistent with Eq.~(\ref{eq:good_cutoff}), and does not
suppress back-scattering.  Using this latter cut-off scheme and $a_0 =
4\;\AA$ we find a magnetic anisotropy more than one order of
magnitude larger than was reported in Ref.~[\onlinecite{brey}]. Furthermore,
with our cut-off scheme the relative value of the anisotropy is rather
insensitive to the precise value of $a_0$.

\begin{figure}
\epsfxsize=2.5in
\begin{center}
\includegraphics[width=8.5cm,clip]{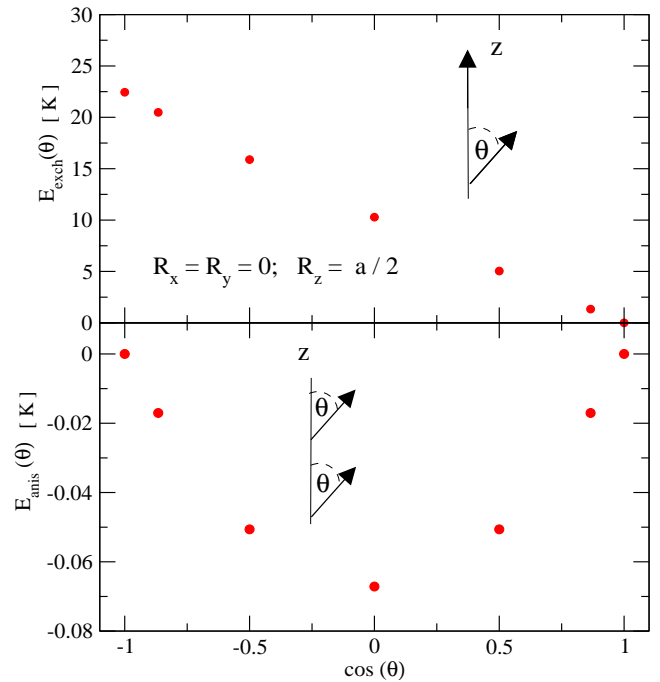}
\end{center}
\vspace*{0.05cm} \caption{\label{fig:angledep} Spin orientation
dependence of the ground state energy for a fixed number of holes
corresponding to a hole concentration of $p = 0.2 {\rm nm}^{-3}$:
(a) One of the spins is aligned along the $z$-axis, and the other
is rotated in the $xz$ plane. (b) The spins are parallel and
rotated in the $xz$ plane.}
\end{figure}

In Fig.~\ref{fig:mesh_dep} we show the obtained effective
interaction as a function of the hole concentration $p$ for two
different numbers of $k$-points ($N$) below the cut-off energy
$E_{\rm cutoff} = 3273 {\rm K}$ for a Mn separation of
$R=a/2\approx 2.8 \AA$.  For these calculations we used a
dimensionless coupling, $j_{pd} \equiv {S J_{pd} m\over a^3 4
\pi^3} =0.2$, with $S=5/2$ the Mn spin, $m$ the free electron
mass, and $a=5.65 \AA$ the lattice constant. 
This value of $j_{pd}$ roughly corresponds
to the exchange coupling used in Ref.~[\onlinecite{brey}]. The
accuracy of the calculation can be inferred both from the
difference between the $N = 3220$ and $N = 1688$ data, and from
the size of the discrete jumps. It is around $10-20 \%$ at best.
Thus, one cannot give a quantitatively reliable estimation of the
anisotropy from these numerical calculations.  In this regard it
is somewhat misleading that for a {\em fixed} number of holes the
exchange energy behaves very nicely as a function of the angle of
the two magnetic impurities,\cite{brey} as demonstrated in
Fig.~\ref{fig:angledep}. In fact, our experience with other local
density of states calculations shows that one may need around
$10^6$ k-points to achieve an accuracy of $\sim 1 \%$, which seems
to be beyond the scope of present-day numerics.\cite{footnote}

It is therefore questionable how well these data can be trusted in
regard to the estimation of the anisotropy energy. However, a
qualitative assessment can be made. While anisotropy energies
$\delta J$ are definitely smaller than the numerical errors of the
total exchange energies, they still show a consistent pattern (see
Fig.~\ref{fig:mesh_dep}), suggesting that the {\em order of
magnitude} of the anisotropy $\delta J / J_{||}$ is probably
correctly given, even for these small values of $N$.

\begin{figure}
\epsfxsize=2.5in
\begin{center}
\includegraphics[width=8.5cm,clip]{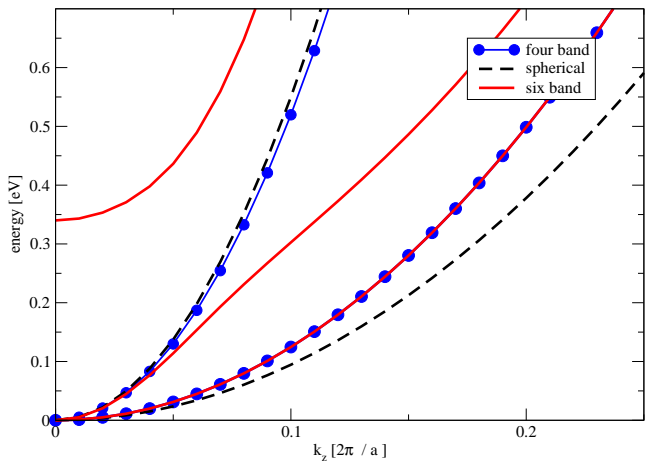}
\end{center}
\vspace*{0.05cm} \caption{\label{fig:bands} Comparison of the hole
dispersion within the full six band model, the four band model,
and the spherical approximation. The latter two are reasonable
approximations for $E_F < 0.15-0.2 {\rm eV}$, which corresponds to
a hole density of $p < 0.25 -  0.4 {\rm nm}^{-3}$. }
\end{figure}

The anisotropy in Fig.~\ref{fig:mesh_dep} for a Mn separation
of $R\approx 2.8\AA$ is  much smaller than the
one obtained within the spherical approximation. This somewhat surprising
result can be understood  as follows:  for short Mn-Mn separations the
exchange interaction is dominated by high energy electron-hole excitations.
However, as shown in Fig.~\ref{fig:bands},  the four band model
($\Delta_{SO} \to \infty$) provides a good approximation to the
exact eigenstates only up to energies  $E \sim 0.15 - 0.2 {\rm eV}$,
corresponding to hole concentrations up to $p\sim 0.3 - 0.4{\rm
nm}^{-3}$.  For these high-energy excitations it therefore fails
and largely overestimates the strength of spin-orbit interaction.

On the other hand, for hole concentrations less than $p\sim 0.3 \
{\rm nm}^{-3}$ the Fermi energy is in a range where the
$\Delta_{SO}\to \infty$ approximation for the single particle
states is appropriate. While for short Mn-Mn separations,
electron-hole excitations at all energy scales contribute to the
RKKY interaction, for larger values of $R$ the RKKY interaction is
dominated by electron-hole excitations in the vicinity $\delta k
\sim 1/R$ of the Fermi surface.  We expect therefore that the
anisotropy will {\em increase} for these concentrations with
increasing Mn-Mn separation $R$.

\begin{figure}
\epsfxsize=2.5in
\begin{center}
\includegraphics[width=8.5cm,clip]{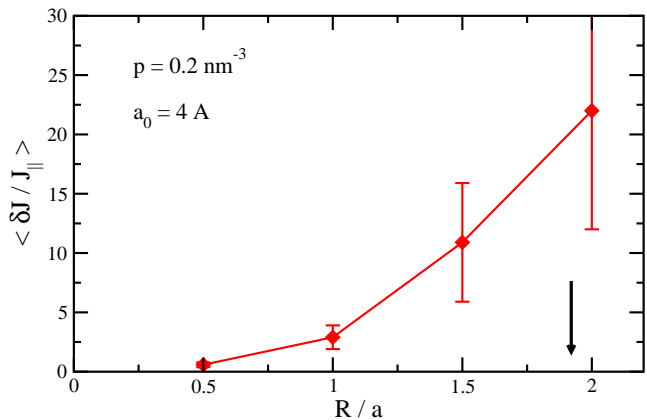}
\end{center}
\vspace*{0.05cm} \caption{\label{fig:anisotropy_of_R} Average
anisotropy for $0.2 {\rm nm}^{-3} < p < 0.3 {\rm nm}^{-3}$. We
used the same parameters as in Fig.~\ref{fig:mesh_dep}. The
anisotropy increases with distance and becomes of the size $\sim
20\%$ for $R=2a$, which is around  the typical Mn-Mn distance for
$x=5 \%$, indicated by the arrow. }
\end{figure}

These expectations are indeed supported by our numerical results shown in
Fig.~\ref{fig:anisotropy_of_R}, where we  find that the
typical anisotropy, $\delta J/J_{||}$ increases with the Mn-Mn
separation and becomes of the order of $\sim 20 \%$ for the
typical Mn-Mn separation $R\sim 10 \AA$  at $x=0.05$ Mn content.

\begin{figure}
\epsfxsize=2.5in
\begin{center}
\includegraphics[width=8.5cm,clip]{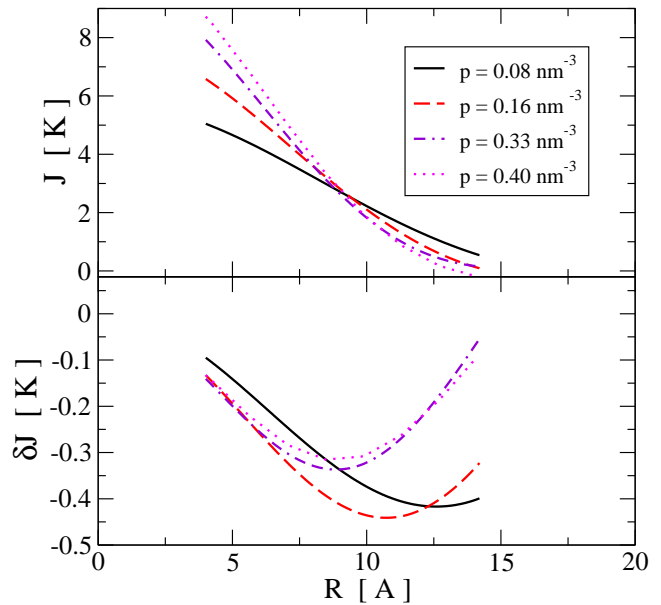}
\end{center}
\vspace*{0.05cm} \caption{\label{fig:J_of_R}
Position- and hole concentration-dependence of the
exchange and anisotropy of the RKKY interaction within the six band model.
}
\end{figure}

In Fig.~\ref{fig:J_of_R} we show the exchange interaction and the
anisotropy as a function of the separation of the two Mn ions for
various hole concentrations. The exchange interaction increases
monotonically as the hole concentration is increased, while the
``interaction range'' decreases due to the increase of the Fermi
momentum. The maximum of the anisotropy term, on the other hand,
decreases for larger hole concentrations, since then the Fermi energy
moves into a range where spin-orbit coupling is of lesser importance.
Depending on the specific value of hole concentration, the size of the
anisotropy is in the range of $10-30 \%$ for typical Mn separations.

Regarding the directional anisotropy (from the cubic symmetry of the
lattice) mentioned in Ref.~[\onlinecite{brey}] we remark that while
spin anisotropy can induce a frustrated ground state and may thus also
change the universality class of the ferromagnetic transition,
directional anisotropy only increases the already existing disorder
somewhat, and does not induce any frustration. As a consequence, it
does not change the magnetic properties of the system, and plays an
unimportant role in an already disordered ferromagnet.

In the calculations above we neglected disordered Coulomb
 scattering of the valence holes on static impurities (As antisites
and Mn core charges {\em etc.}).  This type of
disorder destroys the coherent propagation of the electrons and therefore
reduces the value of both $J_{||}$ and $\delta J$ for separations
larger than the electronic mean free path $\ell$. For the metallic
samples $\ell$ is of the order of the typical Mn-Mn separation, and
therefore we expect that the suppression is not dramatic. While
it is not clear how much the ratio $\delta J/ J_{||}$ is reduced
by static disorder, it is likely that random anisotropy effects are
important for the disordered samples too.

\section{Monte Carlo Study}
\label{sec:montecarlo}

Having obtained the effective kernels, in this section we report the results
of classical Monte Carlo (MC) calculations in the spin degrees of
freedom used to study the implications of anisotropy on the magnetic
properties of ${\rm Ga}_{1-x}{\rm Mn}_x{\rm As}$.  Similar Monte Carlo
studies of GaMnAs have also been carried out on other
models\cite{schliemann_MC,kennett,brey}.

The properties of the obtained ferromagnetic state are expected to
depend on correlations between the positions of Mn spins. As we
mentioned in the introduction, it is presently unknown from the
available experimental data to what extent the substitutional,
magnetically active, Mn positions are correlated. It is therefore
worthwhile to investigate theoretically the trends in magnetic
properties as a function of increasing correlations in the Mn ion
positions.  We applied the spin-spin interaction kernels computed
in the first part of this paper to monitor how the magnetic
properties such as the saturation magnetization, the Curie
temperature, the shape of the magnetization curves and the spin
distributions change as Mn positional correlations are gradually
built in via the procedure described below.

To simulate positional correlations and to control disorder we
first generate a completely random Mn distribution within the Ga
sublattice of an $L\times L\times L$ cube of ${\rm Ga}_{1-x}{\rm
Mn}_x{\rm As}$, and then relax the Mn ions through a standard $T=0$
Monte Carlo procedure with nearest and next nearest neighbor hopping,
assuming a screened Coulomb interaction between the Mn ions.  It is
extremely important to use periodic boundary conditions in the course of
this relaxation procedure, otherwise the Mn ions accumulate on the
surface of the cube.

\begin{figure}[h]
\begin{center}
\epsfig{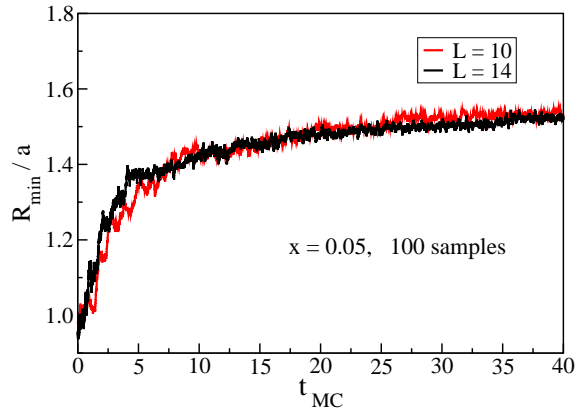}
\caption{
Average nearest neighbor Mn distance   vs. Monte Carlo time
for $L=12 a$ and $L=17 a$.
\label{fig:relax_time}}
\end{center}
\end{figure}

The typical evolution of the average nearest neighbor distance
with MC time, $t_{MC}$, is shown in Fig.~\ref{fig:relax_time} for
two different sample sizes and $x=0.05$. Correlations are formed
within a MC time span of $t_{MC}\sim 5$.  A careful investigation
of these correlations for dilute samples with $x\sim 0.01\;-
\;0.02$ shows that the Mn ions tend to form a somewhat distorted
BCC lattice with point defects.  In the following, we shall use
this Monte Carlo time as a parameter to control the amount of
positional disorder.

In all simulations to be described below the Mn spins were replaced by
their classical angular variables ${\bf S \to} S\;{\bf \Omega}$, which
is expected to be a reasonable approximation for $S=5/2$.  To take
account of the finite mean free path $l \approx 7$\AA\ of the valence
holes\cite{ohno_mfp,Priour:prl04}, we used an exponential cutoff for
the RKKY interaction: $K_{\rm para/perp}({\bf R})\to K_{\rm
para/perp}({\bf R}) e^{-R/l}$. We also introduced a hard cutoff for
the effective interaction, $l_{\rm hard} = 2 r_{\rm Mn}$, where
$r_{\rm Mn}$ is related to the Mn concentration, $n_{\rm Mn} = [4\pi
\;r_{\rm Mn}^3/3]^{-1}$. This hard cut-off has been defined to take
into account that the RKKY approximation does not make any sense
beyond the first ``shell'' of neighbors, since Mn ions are very strong
potential scatterers.

\subsection{Monte Carlo results within the spherical approximation}

\begin{figure}[b]
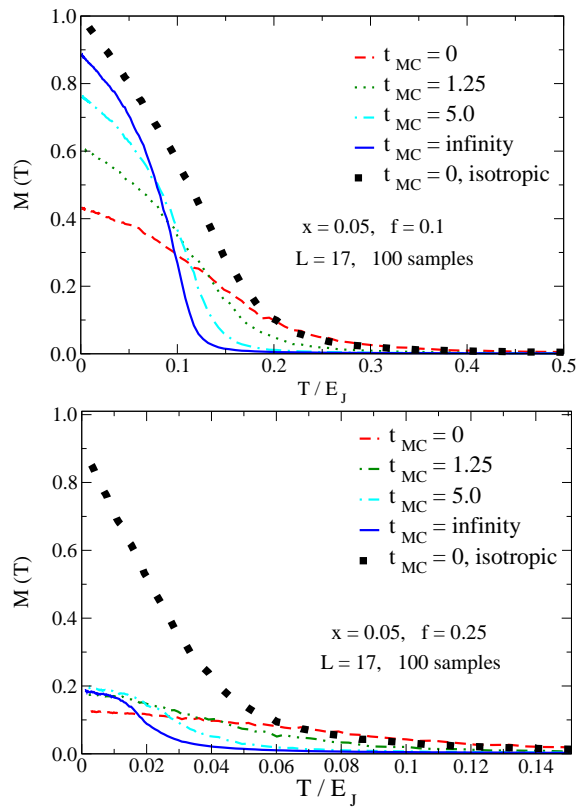

\begin{center}
\epsfig{figure=L17_magnetization_h10_mod.eps, clip, width=7.5 cm}
\epsfig{figure=L17_magnetization_h25_mod.eps, clip, width=7.5 cm}
\caption{
Magnetization vs. temperature
for hole fractions $f=0.10$ (top) and  $f=0.25$ (bottom)
as a function of Monte Carlo time.   Note that for $f=0.1$ 
the isotropic kernel results in nearly full polarization even for the completely disordered sample
demonstrating the the reduction of the magnetization is dominantly due to
anisotropy  effects.
\label{fig:magnetization}}
\end{center}
\end{figure}

Within the spherical approximation, the results depend almost
exclusively on the hole fraction $f$ and do not depend too much on the
specific value of the Mn concentration $x$. This is because the
effective kernel depends on the Mn-Mn distance through $y = k_{F,h}
R$. The typical values of $y$ depend on $f$. However,
lattice-specific effects play a less important role in the dilute
limit, where characteristic distances are typically much larger than
the lattice constant.


In Fig.~\ref{fig:magnetization} we show the temperature-dependence
of the magnetization $M\equiv \vert \langle \Omega_i \rangle
\vert$, for different amounts of disorder, as a function of
temperature for two different hole fractions, $f=0.1$ and
$f=0.25$.  A spontaneous magnetization develops at low
temperatures in both cases.  For small Monte Carlo times (large
disorder) the transition between the paramagnetic and magnetic
phases takes place rather smoothly, and then the magnetization
increases approximately linearly with decreasing temperature,
qualitatively similar to many experiments.\cite{vanesch,potashnik}
The Curie temperature (estimated by where the curves would
intersect the temperature axis if the high temperature tails are
ignored) decreases with {\it decreasing} disorder, {\em i.e.},
disorder tends to {\it enhance} the transition
temperature.\cite{berciu,bhatt,Berciu:prb04} While the $M(T)$
curves of the unrelaxed samples do not quite look like usual
mean-field magnetization curves, they become more and more
mean-field-like upon relaxing the Mn impurity positions.  All
these properties are characteristic of strongly disordered magnets
and have been reported earlier.\cite{berciu}


For both hole fractions we find that the magnetization tends to a
value at $T=0$ that is {\em smaller} than that of a fully polarized
ferromagnet. This effect is mostly due to anisotropy induced
frustration: We recover the fully polarized state when we substitute
the kernel with its angle averaged value.

\begin{figure}[h]
\begin{center}
\epsfig{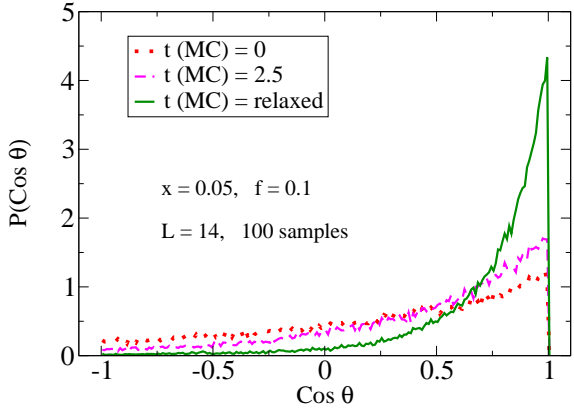}
\caption{Ground state spin distribution for $f=0.10$.  The angle $\theta$ is
measured with respect to the ground state magnetization.
\label{fig:spin_dist_h10}}
\end{center}
\end{figure}

\begin{figure}[h]
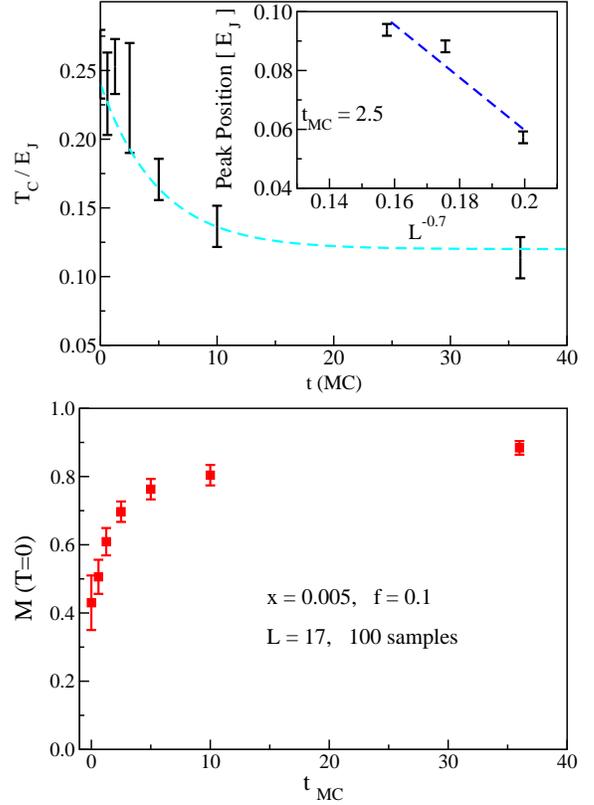

\begin{center}
\epsfig{figure=Curie_tmp_t_MC_h10_mod.eps,clip, width=7.5 cm}
\epsfig{figure=M_L17_0_h10_mod.eps,clip, width=7.5 cm }
\caption{{ Top panel}: the $L=\infty$ extrapolated Curie
temperature as a function of Monte Carlo time.  The dashed curves
are a guide to the eye; {Inset to top panel}: The scaling of
the peak in susceptibility with system size $L$, for $t_{MC} =
2.5$;{Bottom panel}: $M(T=0)$ for $f=0.10$ and $L=17 a$.
\label{fig:L17_mag_and_Tc_vs_tmc_f10} }
\end{center}
\end{figure}

Correlations between the Mn sites decrease frustration effects
and tend to increase this remnant magnetization,
and for $f=0.1$ the fully relaxed system recovers $90 \%$ of
the magnetization (see Fig.~\ref{fig:L17_mag_and_Tc_vs_tmc_f10}).
The corresponding evolution of the distribution of the
ground state spin orientations is shown in Fig.~\ref{fig:spin_dist_h10}.
The angle $\theta$ in Fig.~\ref{fig:spin_dist_h10}
denotes the angle with respect to the  direction of
the  ground state magnetization vector, ${\bf n}: {\rm cos}\theta_i={\bf
\Omega_i \cdot n}$.  When all spins are aligned, $P({\rm
cos}\theta)=\delta(1-{\rm cos}\theta)$, and the more ordered the
positions of the Mn ions, the more peaked the distribution becomes
around $\cos\theta = 1$.

The simulations with $f=0.25$ show a much stronger reduction of
the zero temperature magnetization relative to the case of
$f=0.10$, even for the relaxed samples.  Furthermore, for a fixed
disorder the Curie temperature is reduced by about $20\;-\;40\%$
for $f=0.25$ with respect to that of $f=0.1$, even if we take into
account the factor of $\sim 3$ increase in the energy scale $E_J$.
In view of the results of the following section, we believe that
these results are artifacts of the spherical approximation: As we
emphasized earlier, within the spherical approximation
oscillations appear in the interaction kernel, which show rather
specific features around $y = k_{F,h} R\sim 2$, which is just the
typical value of $y$ for $f=0.25$.  While for $f=0.10$ the
parallel component of the kernel is ferromagnetic for typical
Mn-Mn separations, for $f=0.25$ it becomes antiferromagnetic.
Therefore it is likely that for $f=0.25$ both the {\em anisotropy}
and the {\em antiferromagnetic} part of the RKKY coupling play an
important role.\cite{Zhou:prb04}

The main effects due to correlations between the Mn impurities are
summarized in Fig.~\ref{fig:L17_mag_and_Tc_vs_tmc_f10}, where we show
the $T=0$ magnetization for $L/a=17$ and the $L=\infty$ extrapolated
value of $T_C$ as a function of Monte Carlo time.  This latter has
been obtained by measuring the maximum of the susceptibility for
various system sizes and then extrapolating to $L=\infty$ by using the
critical exponents for the Heisenberg model known from $\epsilon$
expansion.\cite{Guillou:prb80} Both quantities change monotonically with
disorder, with a time scale similar to the one with which the disorder
changes (see Fig.~\ref{fig:relax_time}).

\subsection{Monte Carlo results within the six band model}

In this subsection we study the magnetic properties of
Ga$_{1-x}$Mn$_x$As using the effective interaction kernel computed
within the six band model.  In this case the interaction kernel
depends on the specific direction of the two Mn ions, and, in
principle, one should compute it for {\em all} possible positions
of the two Mn ions, as the effective interaction is, in general, a
rather complicated tensor.  This is next to impossible, and
besides, we do not expect to obtain {\em quantitatively} correct
results from it anyway, since in these calculations we neglect the
polarization of the holes in the ferromagnetic state, the strong
potential scattering off the Mn cores, and the effects of various
defects in Ga$_{1-x}$Mn$_x$As.

Instead, to obtain a qualitative picture, we will pursue the
following strategy: By tetragonal symmetry, the effective Mn spin-spin
interaction kernel reduces to the simple form Eq.~(\ref{eq:Heff_2})
provided that the two Mn ions are aligned along the $x$, $y$, or $z$
direction.  We will therefore use the effective interaction
Eq.~(\ref{eq:Heff_2}), but we will substitute the kernels in this
expression by the ones we computed in Section~\ref{sec:six_band}. In
this way we obtain a qualitatively correct description of the
interaction between the Mn spins in Ga$_{1-x}$Mn$_x$As, which captures
approximately the spin-orbit coupling induced anisotropy effects.

In this case the structure of the $M(T)$ curves depends not only
on the hole fraction $f$, but also on the Mn concentration. In
fact, our results show that for the active Mn concentration range
$0.03 < x < 0.05$, $T_C$ is very sensitive to the Mn concentration
$x$, but exhibits a much weaker dependence on the hole fraction.
This originates from the specific property of the interaction
kernel shown in Fig.~\ref{fig:J_of_R}. As we showed in the
previous section, the role of spin-orbit coupling induced
anisotropy is also more pronounced for larger Mn-Mn separations,
{\em i.e.}, smaller Mn concentrations.

\begin{figure}[h]
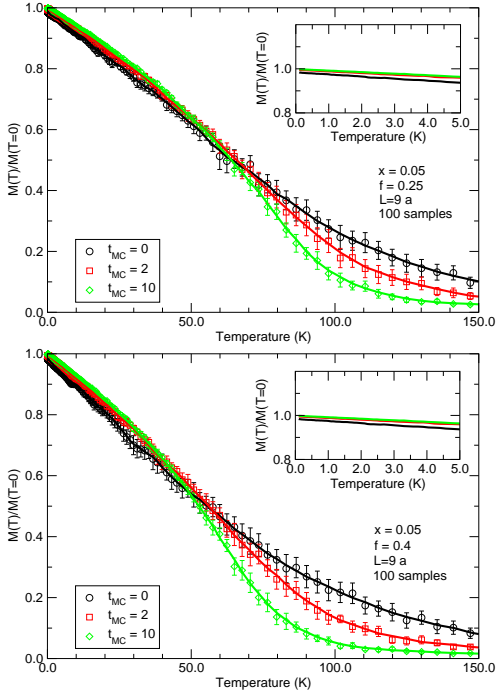

\begin{center}
\epsfig{figure=M_of_T_sixb_x0.05_f0.25_L9_mct_0_2_10.eps, clip,width=6.5 cm}
\epsfig{figure=M_of_T_sixb_x0.05_f0.4_L9_mct_0_2_10.eps, clip,width=6.5 cm}
\caption{Temperature-dependence
 of the magnetization for a sample of linear size
$L=9\;a$ with active Mn concentration $x=0.05$, hole fractions $f=0.25$ and
$f=0.4$ as a function of Monte Carlo relaxation time. The insets show the low temperature
saturation of the magnetization.}
\label{fig:M_of_T_sixb_x=0.05}
\end{center}
\end{figure}

\begin{figure}[h]
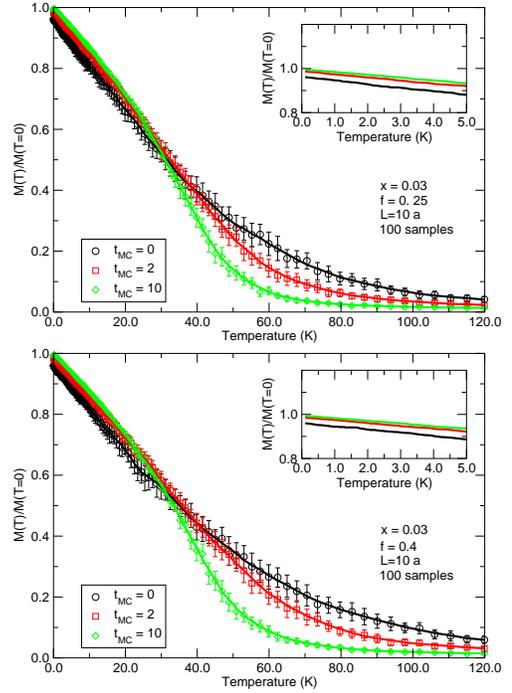

\begin{center}
\epsfig{figure=M_of_T_sixb_x0.03_f0.25_L10_mct_0_2_10.eps, clip,width=6.5 cm}
\epsfig{figure=M_of_T_sixb_x0.03_f0.4_L10_mct_0_2_10.eps, clip,width=6.5 cm}
\caption{Temperature-dependence
 of the magnetization for a sample of linear size
$L=10\;a$ with active Mn concentration $x=0.03$, hole fractions $f=0.25$ and
$f=0.4$ as a function of Monte Carlo relaxation time. The insets show the low temperature
saturation of the magnetization.}
\label{fig:M_of_T_sixb_x=0.03}
\end{center}
\end{figure}

Typical magnetization curves are shown in
Figs.~\ref{fig:M_of_T_sixb_x=0.05} and
\ref{fig:M_of_T_sixb_x=0.03} for $x=0.05$ and $x=0.03$.  For
$x=0.05$ the ground state is always almost fully polarized, even
for the fully disordered, unrelaxed sample, only $T_C$ is
gradually suppressed upon relaxing the impurity positions.  For
$x=0.03$, however, a slight non-collinearity appears for the fully
disordered sample, corresponding to a $4\%$ suppression of the
total magnetization and a typical angle between the ground state
of the sample and an individual spin of the order of $\theta\sim
16$ degrees.  This non-collinearity disappears once we introduce
correlations between the Mn sites.  These results clearly show
that for larger Mn concentrations the spherical approximation
badly fails and a more complete six band model must be used.

We also find that for even smaller concentrations, $x< 0.03$, the
spin-orbit coupling induced disorder plays a more important role, and the
obtained ground state is non-collinear, similar to the one obtained
within the spherical approximation.  These results indicate that
non-collinear states may appear for smaller Mn concentrations. This
small concentration regime is, however, definitely out of reach for
an RKKY approach: At these small concentration disorder also plays an
important role and most likely an impurity band description of the
material is necessary\cite{fiete,berciu,Berciu:prb04}.

\section{Conclusions}
\label{sec:conclusions}

In this paper we continued the route of Ref.~[\onlinecite{zarand}]
to build effective spin models for metallic ${\rm Ga}_{1-x}{\rm
Mn}_x{\rm As}$. We constructed interactions between the Mn spins
mediated via indirect exchange by using two different valence band
models.  First we computed the effective interaction within the
spherical approximation, where the calculations can be performed
analytically, and then we studied the effective interaction
numerically using the more complete six band model, also studied
in Ref.~[\onlinecite{brey}].

We find, in agreement with Ref.~[\onlinecite{brey}],
that the spherical approximation badly fails for short Mn
separations where it overestimates the strength of the spin-orbit
coupling induced anisotropy
by more than one order of magnitude. However, we find using
the six band model, that the strength of the spin-orbit coupling
induced anisotropy {\em
increases} with Mn-Mn distance, and becomes of the order of $\sim
20\%$, qualitatively (but not quantitatively) agreeing with the
results obtained using the spherical approximation and also in rough
agreement with the experimentally observed FMR
linewidth.\cite{Rappoport:prb04,Liu:prb03} This implies that the
frustration effects discussed in Ref.~[\onlinecite{zarand}] are
probably more pronounced for small Mn concentrations.

We carried out classical Monte Carlo simulations to study the
implications of the computed effective interactions.  Within the
spherical approximation, we always find a generically non-collinear
state due to orientational frustration (random anisotropy). It has
been speculated earlier that this state may be a ferromagnetic state
of spin glass nature.\cite{zarand} However, our studies of hysteresis
indicate that this random anisotropy results in a conventional, though
non-collinear and disordered ferromagnetic state.  This result seems
to be supported by the fact that random anisotropy is presumably
irrelevant at the Heisenberg fixed point,\cite{Chayes:prl86} and
therefore does not change the critical scaling of a Heisenberg magnet.

The strongly non-collinear states found within the spherical model are
partly an artifact of the spherical approximation, which is only valid
at sufficiently small hole concentrations. Using the effective
interaction obtained within the six band model of the valence holes,
we find that the anisotropy effects are too weak to induce a strongly
non-collinear state for (active) Mn concentrations as large as
$x=0.05$.  However, anisotropy becomes more important for lower
concentrations, where it may induce a frustrated and non-collinear
state.\cite{fiete} In particular, for $x=0.03$ we find that the ground
state of the fully disordered sample is not fully collinear, and
individual spins deviate from the ferromagnetic orientation of the
sample by about $\sim 16$ degrees.  This tendency is expected to be
even more pronounced for lower Mn concentrations where we expect a
non-collinear ferromagnetic state.\cite{fiete} Indeed, for Mn
concentrations in the range $x=0.015 - 0.02$ we find a non-collinear
state similar to the one obtained in the spherical approximation,
though this concentration range may already be well out of the range
of the metallic approximation used in this paper, and the impurity
band approach of Refs.~[\onlinecite{fiete,berciu,Berciu:prb04}] may be
more appropriate in this limit. Experimental evidence also supports
the presence of a non-collinear state in the regime of small 
Mn concentrations.\cite{Potashnik:apl03}

Let us emphasize again, that the  Mn concentrations above denote the
concentration of {\em active} substitutional Mn ions, which can be
substantially smaller than the nominal Mn concentration of the
samples.\cite{yu,edmonds} In particular, it is known that a
substantial amount of the dopant Mn ions go to interstitial sites
in course of the out-of equilibrium growth
process,\cite{yu,edmonds} and these interstitial Mn ions are also
believed to bind to the substitutional Mn ions and ``neutralize''
them from the point of view of ferromagnetism.\cite{bergqvist} In
this way, it is quite possible that an unannealed sample with a
nominal Mn concentration of $x_{\rm nom} \sim 0.05$ has an active
Mn concentration of only around $x = 0.01 - 0.03$ and have a
non-collinear ground state. It is, on the other hand quite likely
that annealed samples and samples with higher Mn concentrations
form a ferromagnetic state where the Mn spins are almost perfectly
collinear.

As first pointed out in Ref.~[\onlinecite{zarand}], the observed
saturation magnetization of ${\rm Ga}_{1-x}{\rm Mn}_x{\rm As}$ is
much less even in annealed samples than expected based on the
nominal Mn concentration of the samples.  The orientational
frustration discussed in this paper provides a possible mechanism
to explain the missing magnetization in unannealed and
``underdoped'' samples, where one can substantially increase the
saturation value of the magnetization, $M(T=0)$ with a relatively
small magnetic field compared to $T_C$, $H< 1\;{\rm
T}$.\cite{Potashnik:apl03} Other mechanisms have also been
proposed: The main mechanism that reduces the magnetization of
annealed samples seems to be provided by the presence of
interstitial Mn ions, which - as mentioned earlier -
``neutralize'' substitutional Mn impurities.\cite{bergqvist} It
has also been proposed that As antisite defects can induce a
non-collinear state,\cite{Korzhavi:prl02} and an intrinsic
non-collinearity has been proposed also in
Ref.~[\onlinecite{schliemann,Schliemann:prl02}]. Finally, quantum
fluctuations due to the antiferromagnetic coupling between the Mn
spins and the valence holes have been also proposed
recently.\cite{MacDonald_private} It is probably a combination of
all these effects that is finally responsible for the missing
magnetization of ${\rm Ga}_{1-x}{\rm Mn}_x{\rm As}$.

Finally, we comment on the approximations made in this paper.
Throughout, we followed a common practice in the
literature,\cite{Timm:rev} and neglected the potential scattering off
the Mn ions. This approximation is expected to produce {\em
qualitatively} reliable results in the metallic limit considered and
we believe that all the trends reported here are robust.  We also
neglected the polarization of the valence holes induced by the
ferromagnetically alligned Mn spins, which may be important at low
temperatures. In summary, we believe the answers to the questions of
how anisotropic the Mn spin-spin interactions are in metallic GaMnAs
and how anisotropy interplays with Mn positional correlations to affect
important magnetic properties are now on more solid ground,
though further studies are needed to understand frustration effects 
in the regime $x<0.03$.

{\em NOTE:} While preparing this work for publication, a preprint
appeared\cite{Timm:condmat04} where the authors arrive at rather
similar conclusions using a very different, tight binding approach.

\begin{acknowledgments}

We would like to thank L. Brey, D. S. Fisher, J. K. Furdyna, X.
Liu, C. Timm, E. Sasaki and F. von Oppen for stimulating
discussions. This work was supported by NSF PHY-9907949,
DMR-0233773, NSF-NIRT-DMR 0210519, NSF-MTA-OTKA Grant No.
INT-0130446, Hungarian Grants No. OTKA T038162, T046267, and
T046303, the European ``Spintronics'' RTN HPRN-CT-2002-00302 and
the Packard foundation. G.Z. has been supported by the Bolyai
Foundation, whereas B. J. would like to gratefully acknowledge
support of the Alfred P. Sloan Foundation.

\end{acknowledgments}

\appendix
\section{Calculation of RKKY Kernel}
\label{app:RKKY_calculation}
In this appendix, we derive explicit expressions for the kernels
$K_{\rm par}(R)$ and $K_{\rm perp}(R)$
appearing in Eq.~(\ref{eq:Heff}). By spherical symmetry, we can assume that the
two Mn impurities are alligned along the $z$ axis.
In this case, the kernel $K^{\alpha\beta}$ becomes diagonal, $K^{\alpha\beta}= K^{\alpha} \delta_{\alpha\beta}$,
and can be written by converting the sums in Eq.~(\ref{eq:RKKY_sum}) to integrals and
assuming a parabolic dispersion for the light and heavy hole bands
as
\begin{widetext}
\begin{equation}
K^{\alpha \alpha}({\bf R})=2G^2 \sum_{\mu \mu'}
\int_0^{k_{F, \mu}} {k^2 dk \over 2 \pi^2}
\int_{k'_{F, \mu}}^{\infty} {k'^2 dk' \over 2 \pi^2}
{1 \over {k'^2 \over 2 m_{\mu'}}-{k^2 \over 2 m_{\mu}}}
\langle |J^{\alpha}_{\mu \mu'}(\theta, \theta',\phi,\phi')|^2
\,2\, {\rm cos}(k R {\rm cos}(\theta)-k' R {\rm cos}(\theta'))
\rangle_{\theta, \theta',\phi,\phi'}\;.
\end{equation}
\end{widetext}
Here we have taken the angles $\theta$ and $\phi$ ($\theta'$ and $\phi'$)
parametrize the directions  $\hat {\bf k} = (\sin\theta \cos \phi,\;\sin\theta \sin \phi,\; \cos\theta)$
and $\hat {\bf k'}$, and the brackets denote the
angular average over the angles $\theta, \theta',\phi$ and $\phi'$.
Making the substitutions $k \to \sqrt{2 m_{\mu} \epsilon_F} k$ and
$k' \to \sqrt{2 m_{\mu'} \epsilon_F} q$ yields the expression
\begin{eqnarray}
K^{\alpha}({\bf R})=8\,\epsilon_F \sum_{\mu, \mu'} g_{\mu} g_{\mu'}
\int_0^1 dk \int_1^{\infty}dq {q^2 k^2 \over q^2-k^2} \nonumber \\
\times I^{\alpha}_{\mu \mu'}(k_{F , \mu} k R, k_{F , \mu'} q R)\;,\nonumber \\
\label{eq:RKKY_integral}
\end{eqnarray}
where
\begin{eqnarray}
I^{\alpha}_{\mu \mu'}(k_{F , \mu} k R, k_{F , \mu'} q R)&=&
\langle |J^{\alpha}_{\mu \mu'}(\theta, \theta',\phi,\phi')|^2
\,\nonumber \\
\times 2\, {\rm cos}(k_{F , \mu} k R {\rm cos}(\theta)&-&
k_{F , \mu'} q R {\rm cos}(\theta'))\rangle_{\theta, \theta',\phi,\phi'}\;,\nonumber \\
\label{eq:I}
\end{eqnarray}
and we introduced the dimensionless couplings, $g_{\mu}=G \varrho_{\mu}=G {m_{\mu}^{3/2}
\over \sqrt{2} \pi^2}\epsilon_F^{1/2}$.

To compute $K^{\alpha \alpha}({\bf R})$ we first evaluate
$I^{\alpha}_{\mu \mu'}(k_{F , \mu} k R, k_{F , \mu'} q R)$ by
performing the angular integrals. Since $I^{\alpha}_{\mu
\mu'}(k_{F , \mu} k R, k_{F , \mu'} q R)$ only depends on $k$ and
$q$ through the combinations $k_{F , \mu} k R$ and $ k_{F , \mu'}
q R$, it is worth defining the heavy hole contributions to it as:
\begin{equation}
I^{\alpha}_{h h} \equiv I^{\alpha}_{\frac32,\frac32}+I^{\alpha}_{\frac32,-\frac32}
+I^{\alpha}_{-\frac32,\frac32}+I^\alpha_{-\frac32,-\frac32} \;.
\end{equation}
The contributions $I^{\alpha}_{h l}$, $I^{\alpha}_{l h}$, and $I^{\alpha}_{l l }$, and the corresponding contributions
to the kernel, $K^{\alpha}_{h h}$, $K^{\alpha}_{l h}$, $K^{\alpha}_{h l}$, $K^{\alpha}_{ll}$,
can be defined  in an analogous way.

We demonstrate the procedure of computing $K^\alpha$ by the
example of the heavy hole contribution to $K^z_{h h}$.  It is
straightforward to evaluate the angular integrals, and for
$I^{z}_{h h}$ one obtains the following expression:
\begin{eqnarray}
I^{z}_{h h}(r k, r q)=
{9 \over 10} {{\rm sin}(k r) \over k r}{{\rm sin}(q r) \over q r} \nonumber \\
+ {45 \over 2} \Biggl({{\rm cos}(k r)\over k^2 r^2} - {{\rm sin}(k r)\over k^3 r^3} + {3\over 5} {{\rm sin}(k r)\over k r}\Biggr )
\nonumber \\
 \times \Biggl ( {{\rm cos}(q r)\over q^2 r^2} - {{\rm sin}(q r)\over q^3 r^3} + {3\over 5} {{\rm sin}(q r)\over q r}\Biggr )\;,
\end{eqnarray}
which can be rewritten as
\begin{eqnarray}
I^{z}_{h h}(r k, r q)={9 \over 10}\frac{ F(k) F(q)}{kq} + {45 \over 2}\frac{ G(k) G(q)}{kq}\;,
\label{eq:F_and_G}
\end{eqnarray}
where $F$ and $G$ have the obvious definitions.
The heavy hole contribution to $K^z(R)$ thus reads
\begin{eqnarray}
K^{z}_{h,h}({\bf R})
=
8\,\epsilon_F g_{h}^2
\int_0^1 \!\!\!\! dk\int_1^{\infty}\!\!\!\!\!\! dq &&{1 \over q^2-k^2} \Biggl [{9 \over 10} k q F(k) F(q) \nonumber \\
&& + {45 \over 2} k q G(k) G(q) \Biggr ]\;.
\label{K^z_hh}
\end{eqnarray}

As shown in Appendix~\ref{app:appendix_int},  for $F$ and $G$ of the
form that appear in  Eq.~(\ref{eq:RKKY_integral})
the integrals over $q$  in Eq.~(\ref{K^z_hh}) can be carried out to give
\begin{eqnarray}
\int_0^1 &dk&\int_1^{\infty} dq{q k \over q^2-k^2} F(k)F(q) =
\nonumber \\
&=&{\pi \over 2} i \int_0^1 dk k \Biggl [F^+(k)^2-F^-(k)^2
\nonumber \\
 &+& F(k) {\rm lim_{q\to 0} }\left \{{2q^2 \over q^2-k^2} F^+(q)\right \} \Biggr ]\;,
\label{eq:indentity_text}
\end{eqnarray}
where $F^+$ and $F^-$ denote the parts of the function $F= F^+ + F^-$ which are analytical
on the upper and lower half-planes, respectively. Applying this formula, we obtain upon making
the substitution $a=k_{F , h} R$ and $b=k_{F , l} R$
\begin{eqnarray}
K^{z z}_{h,h}({\bf R})=4 \pi\,\epsilon_F g_{h}^2
\int_0^1 dk \Biggl [\frac{9}{20}\frac{ k\,{\rm sin}(2 k a )}{a^2} \nonumber \\
+ \frac{45}{2}
\Biggl (\frac{{\rm cos}(k a)}{a^5 k^2}+ \frac{3{\rm cos}(2 k a)}{5 a^3 }- \frac{3 {\rm cos}(2 k a)}{a^5 k^2}- \frac{{\rm sin}(k a)}{a^6 k^3}\nonumber \\
+ \frac{3 {\rm sin}(k a)}{5 a^4 k} + \frac{{\rm sin}(2 k a)}{2 a^6 k^3} -
\frac{11 {\rm sin}(2 k a)}{10 a^4 k} + \frac{9 {\rm sin}(2 k a)}{50 a^2} \Biggr ) \Biggr ]\nonumber \\
\equiv 4 \pi\,\epsilon_F g_{h}^2 C_{\rm para}^{\rm heavy} (a)\;.\;\;\;\;\;\;\;\;\;\;\;\;\;\;
\label{eq:Kzhh}
\end{eqnarray}
It is also possible to evalute the integral (\ref{eq:Kzhh}), but the resulting formula is very long,
and for practical purposes  it is
simpler to evaluate Eq.~(\ref{eq:Kzhh}) numerically.

The remaining parts of the kernel can be evaluated in a similar way, and
are given below. By symmetry, $K^x = K^y$ for this arrangement, and therefore
only the $x$-component is displayed.
The spatial dependence of the various
parts of the kernel is shown in Fig.~\ref{fig:reduced_kernel}.

\begin{widetext}
\begin{eqnarray}
K^{x x}_{h,h}({\bf R})
=4 \pi \,\epsilon_F g_{h}^2
\int_0^1 dk \frac{9}{2} \Biggl [-\frac{{\rm cos}(k a)}{a^5 k^2}
-\frac{{\rm cos}(2 k a)}{a^3}  + \frac{{\rm cos}(2 k a)}{a^5 k^2}\nonumber
+ \frac{{\rm sin}(k a)}{a^6 k^3}  - \frac{{\rm sin}(k a)}{a^4 k}
-\frac{{\rm sin}(2 k a)}{2 a^6 k^3}
+ \frac{3{\rm sin}(2 k a)}{2 a^4 k}\Biggr ]\;.
\end{eqnarray}

\begin{eqnarray}
K^{z z}_{h,l}({\bf R}) &+&K^{z z}_{l,h}({\bf R})= 4 \pi\,\epsilon_F g_{h}g_{l}
\int_0^1 dk
\Biggl [\frac{3}{2} \Biggl (- \frac{{\rm cos}(k a) {\rm cos}(k b)}{a^2 b}
+ \frac{{\rm cos}(k a) {\rm sin}(k a)}{a^6 k^3}
+ \frac{{\rm cos}(k a) {\rm sin}(k a)}{a^4 k}
+\frac{{\rm cos}(k b) {\rm sin}(k a)}{a^3 b k}
\nonumber \\
&-& \frac{{\rm sin}(2 k a)}{2 a^6 k^3}
-\frac{{\rm sin}(2 k a)}{2 a^4 k}
- \frac{{\rm sin}(k b)}{a^3 b k}
+ \frac{{\rm cos}(k a) {\rm sin}(k b)}{a^3 b k}
+\frac{{\rm sin}(k a) {\rm sin}(k b)}{a^2 b}\Biggr )
\nonumber \\
&+&\frac{45}{2} \Biggl (-\frac{{\rm cos}(k a)}{a^2 b^3 k^2}
-\frac{{\rm cos}(k b)}{a^3 b^2 k^2}
+\frac{{\rm cos}(k a) {\rm cos}(k b)}{a^2 b^3 k^2}
+\frac{{\rm cos}(k a) {\rm cos}(k b)}{a^3 b^2 k^2}
+\frac{{\rm sin}(k a)}{a^3 b^3 k^3}
-\frac{{\rm cos}(k b) {\rm sin}(k a)}{a^3 b^3 k^3}
\nonumber \\
&+&\frac{{\rm cos}(k b) {\rm sin}(k a)}{a^2 b^2 k}
+\frac{{\rm sin}(k b)}{a^3 b^3 k^3}
-\frac{{\rm cos}(k a) {\rm sin}(k b)}{a^3 b^3 k^3}
+\frac{{\rm cos}(k a) {\rm sin}(k b)}{a^2 b^2 k}
-\frac{{\rm sin}(k a) {\rm sin}(k b)}{a^2 b^3 k^2}
-\frac{{\rm sin}(k a) {\rm sin}(k b)}{a^3 b^2 k^2}\Biggr)\nonumber \\
&+&\frac{27}{2}\Biggl (- \frac{{\rm cos}(k a) {\rm cos}(k b)}{a b^2}
-\frac{{\rm sin}(k a)}{a b^3 k}
+\frac{{\rm cos}(k b) {\rm sin}(k a)}{a b^3 k}
+\frac{{\rm cos}(k a) {\rm sin}(k b)}{a b^3 k}
+\frac{{\rm cos}(k b) {\rm sin}(k b)}{b^6 k^3}
\nonumber \\
&+&\frac{{\rm cos}(k b) {\rm sin}(k b)}{b^4 k}
+\frac{{\rm sin}(k a) {\rm sin}(k b)}{a b^2}
-\frac{{\rm sin}(2 k b)}{2 b^6 k^3}
-\frac{{\rm sin}(2 k b)}{2 b^4 k}\Biggr ) \Biggr ]\;.
\end{eqnarray}

\begin{eqnarray}
K^{xx}_{h,l}({\bf R})&+ & K^{xx}_{l,h}({\bf R}) =
4 \pi \,\epsilon_F g_{h}g_{l}
\int_0^1 dk
\Biggl [\frac{3}{2} \Biggl (- \frac{{\rm cos}(k a) {\rm cos}(k b)}{a^2 b}
+ \frac{{\rm cos}(k a) {\rm sin}(k a)}{a^6 k^3}
+\frac{{\rm cos}(k a) {\rm sin}(k a)}{a^4 k}
+\frac{{\rm cos}(k b) {\rm sin}(k a)}{a^3 b k}
\nonumber \\
&&- \frac{{\rm sin}(2 k a)}{2 a^6 k^3}
-\frac{{\rm sin}(2 k a)}{2 a^4 k}
- \frac{{\rm sin}(k b)}{a^3 b k}
+ \frac{{\rm cos}(k a) {\rm sin}(k b)}{a^3 b k}
+\frac{{\rm sin}(k a) {\rm sin}(k b)}{a^2 b}\Biggr )
\nonumber \\
&+&\frac{9}{2} \Biggl (\frac{{\rm cos}(k a)}{a^2 b^3 k^2}
+\frac{{\rm cos}(k b)}{a^3 b^2 k^2}
+\frac{{\rm cos}(k a) {\rm cos}(k b)}{a b^2}
-\frac{{\rm cos}(k a) {\rm cos}(k b)}{a^2 b^3 k^2}
-\frac{{\rm cos}(k a) {\rm cos}(k b)}{a^3 b^2 k^2}
-\frac{{\rm sin}(k a)}{a^3 b^3 k^3}
+ \frac{{\rm sin}(k a)}{a b^3 k}
\nonumber \\
&+&\frac{{\rm cos}(k b) {\rm sin}(k a)}{a^3 b^3 k^3}
-\frac{{\rm cos}(k b) {\rm sin}(k a)}{a b^3 k}
-\frac{{\rm cos}(k b) {\rm sin}(k a)}{a^2 b^2 k}
-\frac{{\rm sin}(k b)}{a^3 b^3 k^3}
+\frac{{\rm cos}(k a) {\rm sin}(k b)}{a^3 b^3 k^3}
-\frac{{\rm cos}(k a) {\rm sin}(k b)}{a b^3 k}
\nonumber \\
&-&\frac{{\rm cos}(k a) {\rm sin}(k b)}{a^2 b^2 k}
-\frac{{\rm cos}(k b) {\rm sin}(k b)}{b^6 k^3}
-\frac{{\rm cos}(k b) {\rm sin}(k b)}{b^4 k}
-\frac{{\rm sin}(k a) {\rm sin}(k b)}{a b^2}
+\frac{{\rm sin}(k a) {\rm sin}(k b)}{a^2 b^3 k^2}
+\frac{{\rm sin}(k a) {\rm sin}(k b)}{a^3 b^2 k^2}
\nonumber \\
&+&\frac{{\rm sin}(2 k b)}{2 b^6 k^3}
+\frac{{\rm sin}(2 k b)}{2 b^4 k} \Biggr )
+3 \Biggl (\frac{k\,{\rm cos}(k a) {\rm sin}(k a)}{a^2}
+\frac{k\,{\rm cos}(k b) {\rm sin}(k a)}{a b}
-\frac{k\,{\rm sin}(2 k a)}{2 a^2}
+\frac{k\,{\rm cos}(k a) {\rm sin}(k b)}{a b}
\nonumber \\
&+&\frac{k\,{\rm cos}(k b) {\rm sin}(k b)}{b^2}
-\frac{k\,{\rm sin}(2 k b)}{2 b^2} \Biggr) \Biggr]\;.
\end{eqnarray}

\begin{eqnarray}
K^{zz}_{l,l}({\bf R})&=&4 \pi \,\epsilon_F g_{l}^2
\int_0^1 dk \Biggl [\frac{9{\rm sin}(2 k b)}{20 a^2}
+\frac{45}{2}\Biggl (\frac{{\rm cos}(k b)}{a^5 k^2}
+\frac{{\rm cos}(2 k b)}{15 b^3} 
-\frac{{\rm cos}(2 k b)}{b^5 k^2}
-\frac{{\rm sin}(k b)}{b^6 k^3}
+\frac{{\rm sin}(k b)}{15 b^4 k}\nonumber \\
&+&\frac{{\rm sin}(2 k b)}{2 b^6 k^3}
-\frac{17{\rm sin}(2 k b)}{30 b^4 k}
+ \frac{k\,{\rm sin}(2 k b)}{450 b^2} \Biggr ) \Biggr]\;.
\end{eqnarray}

\begin{eqnarray}
K^{xx}_{l,l}({\bf R})&=&
4 \pi\,\epsilon_F g_{l}^2
\int_0^1 dk \frac{9}{2}\Biggl [-\frac{{\rm cos}(k b)}{b^5 k^2}
+\frac{{\rm cos}(2 k b)}{3 b^3}
+\frac{{\rm cos}(2 k b)}{b^5 k^2}
+\frac{{\rm sin}(k b)}{b^6 k^3}
+\frac{{\rm sin}(k b)}{3 b^4 k}
-\frac{{\rm sin}(2 k b)}{2 b^6 k^3}
\nonumber\\&&
+\frac{{\rm sin}(2 k b)}{6 b^4 k}
+\frac{4 k\,{\rm sin}(2 k b)}{9 b^2}  \Biggr ]\;.
\end{eqnarray}
\end{widetext}

\section{Evaluation of Singular RKKY Integrals for Spin 3/2 Particles}
\label{app:appendix_int}
In this appendix we establish the identity Eq.~(\ref{eq:indentity_text}).
First we note that the functions $F(k)$ that appear in the evaluation of the RKKY
kernel, Eq.~(\ref{eq:RKKY_integral}), have two important properties
that we will use in course of the derivation:
(i) they are odd, $F(-k)=-F(k)$ and (ii)  $F$ is regular at the origin.


First let us prove that
\begin{equation}
I={\rm lim}_{\delta \to 0}\int_{-1+i\delta}^{1+i\delta} dz \int_{-1}^1 dq {q z \over q^2-z^2} F(z)F(q)=0\;,
\label{eq:integral}
\end{equation}
where $\delta$ is a positive infinitesimal quantity.  Decomposing the integrand
${q  \over q^2-z^2} = \frac{1}{2}\left (\frac{1}{q-z}+\frac{1}{q+z}\right )$, introducing the variable
$x=z-i\delta$ $I$, and using the identity $\frac1{x\pm i \delta} = {\cal P}\frac 1x \mp i \pi \delta(x)$
we rewrite $I$ as
\begin{eqnarray}
I &= &\int_{-1}^1 dx \int_{-1}^1 dq\Biggl [P {q \over q^2-x^2}
\nonumber \\
&+&i \frac{\pi}{2} \delta(q-x)-i \frac{\pi}{2} \delta(q+x) \Biggr ] x F(x)F(q) \;,
\end{eqnarray}
where $P$ denotes the principal part of the integral.  The principal
value integral over the first term is zero by symmetry, and the integral over
the last two terms terms simply evaluates to $i\frac{\pi}{2} (q F(q) F(q)- (-q) F(-q) F(q) )=0$,
since $F(-q)=-F(q)$.

We now return to the formula, Eq.~(\ref{eq:indentity_text}).  Using Eq.~(\ref{eq:integral})  and the fact
that the integrand in Eq.~(\ref{eq:indentity_text}) is even in both $k$ and $q$,
we can extend the region of integration to obtain
\begin{eqnarray}
&&\int_0^1 dk \int_1^{\infty}dq{q k \over q^2-k^2} F(k)F(q)=
\nonumber \\
&&{\rm lim}_{\delta \to 0} \frac{1}{8} \int_{C} dz \int_{-\infty}^\infty dq\; z \;k F(z)
\left (\frac{1}{q-z}+\frac{1}{q+z}\right ) F(q)\;,
\nonumber
\end{eqnarray}
where $C$ denotes the contour $z\in [-1+i\delta, 1 + i\delta]$.

\begin{figure}[t]
\begin{center}
\epsfig{figure=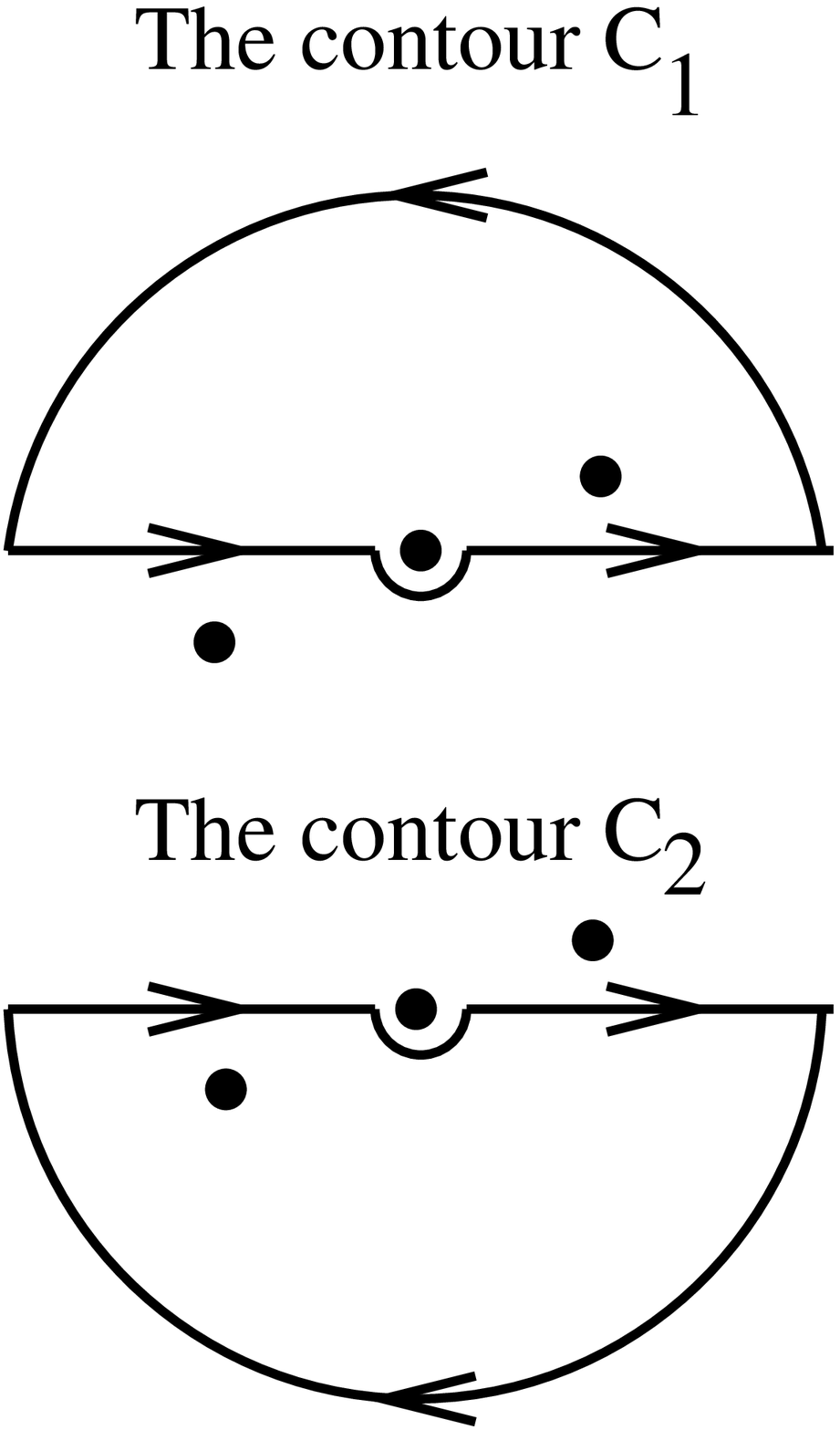, height=3in}
\caption{The contours used in the evaluation of Eq.~(\ref{eq:decomposed}). The dots indicate possible
singularities of the integrand.
\label{fig:contour}}
\end{center}
\end{figure}

We now turn our attention to the integral
\begin{eqnarray}
R \equiv \int_{-\infty}^\infty dq  \left (\frac{1}{q-z}+\frac{1}{q+z}\right ) F(q)\;.
\label{eq:R}
\end{eqnarray}
We first deform the contour of $q$ such that it passes below the origin using the fact that the
integrand of  Eq.~(\ref{eq:R}) is analytical at the origin. Then
 we decompose the function $F$ as $F=F^+ + F^-$, with  $F^+$ and $F^-$ being analytical functions
in the upper and lower half-planes (apart from possible singularities at the origin).
For example, in Eq.~(\ref{eq:F_and_G}), we rewrite  $F(q)$ as $F(q)=
  {{\rm sin}(q r) \over  r} = {e^{iq r} \over 2 i r} - {e^{-iq r} \over 2 i r}$,
and we can similarly decompose $G(q)$ in Eq.~(\ref{eq:F_and_G})
in the same way term  by term.
We can now close the contours for the part $\sim F^+$ on the upper half-plane, and
similarly for $\sim F^-$ on the lower half-plane as shown
in Fig.~\ref{fig:contour} to obtain
\begin{eqnarray}
R=\int_{C_1} \left(\frac{1}{q-z}+\frac{1}{q+z}\right )F^+(q) \nonumber \\
+\int_{C_2} \left(\frac{1}{q-z}+\frac{1}{q+z}\right ) F^-(q)\;,
\label{eq:decomposed}
\end{eqnarray}
with $z= k+i\delta$.
Evaluation of the first integral gives
$2\pi i \left [F^+(k) + {\rm Res_{q\to0}}\left \{{2q^2 \over q^2-k^2} F^+(q)\right \} \right ]$ and
 the second integral gives $2\pi i (-1) F^-(-k)$.  Using the property
$F^-(-k)=-F^+(k)$ we obtain the final result
\begin{equation}
R=2\pi i \left [ 2F^+(k) + {\rm Res_{q\to0}}\left \{{2q \over q^2-k^2} F^+(q)\right \} \right ]\;.
\end{equation}
Since the worst singularity is $F^\pm(q) \sim q^{-2}$,
we can write the residue part of this expression as
\begin{equation}
{\rm Res_{q\to0}}\left \{{2q \over q^2-k^2} F^+(q)\right \}  =
{\rm lim_{q\to0}}\left \{{2q^2 \over q^2-k^2} F^+(q)\right \}\;,
\end{equation}
and we finally obtain
\begin{eqnarray}
&\int_0^1 dk \int_1^{\infty}dq{q k \over q^2-k^2} F(k)F(q)= \\
&\frac{1}{8}
\int_{-1}^1 dk k F(k) 2\pi i \left [ 2F^+(k) + {\rm lim_{q\to0}}\left \{{2q^2 \over q^2-k^2} F^+(q)\right \} \right ]\;.
\nonumber
\end{eqnarray}
In the first term we can make the replacement under the integral
$F^+(k)=\frac{1}{2} \left(F^+(k)+F^+(-k)\right ]$ since $k F(k)$ is even.
Recalling that $F(q)=F^+(q)+F^-(q)$ and $F^+(-k)=-F^-(k)$ we then obtain,
\begin{eqnarray}
&&\int_0^1 dk \int_1^{\infty}\!\!\!\!\!\!dq{q k \over q^2-k^2} F(k)F(q)
\\
&&=\frac{\pi}{2}
\int_0^1 dk k  \Biggl [F^+(k)^2-F^-(k)^2\nonumber \\
&&+ F(k){\rm lim_{q\to0}}\left \{{2q^2 \over q^2-k^2} F^+(q)\right \} \Biggr ]\;,
\end{eqnarray}
which is just the identity we wanted to establish.



\end{document}